\documentclass[aps,pra,superscriptaddress,twocolumn]{revtex4-1}
\usepackage{amsmath,amsfonts,amsthm} 
\usepackage{graphicx}
\usepackage{comment}
 \usepackage{booktabs}
 \usepackage{appendix}
\usepackage{xcolor}
\usepackage{physics}
\usepackage{hyperref}
\usepackage{bm}
\usepackage{chngcntr}
\usepackage[utf8]{inputenc}

\newcommand{\mike}[1]{\textcolor{black}{#1}}

\begin{document}


\title{Dynamical Drexhage Effect: Amplified Emission in Time-Modulated Electromagnetic Environments}

\author{J. C. Obeso Jureidini}
\email{jobesojureidini@ucsd.edu}
\affiliation{Department of Chemistry and Biochemistry, University of California San Diego, La Jolla, California 92093, USA}
\author{M. Reitz}
\email{mireitz@ucsd.edu}
\affiliation{Department of Chemistry and Biochemistry, University of California San Diego, La Jolla, California 92093, USA}
\author{P. Fowler-Wright}
\affiliation{Department of Chemistry and Biochemistry, University of California San Diego, La Jolla, California 92093, USA}
\author{A. Koner}
\affiliation{Department of Chemistry and Biochemistry, University of California San Diego, La Jolla, California 92093, USA}
\author{H. Nasari}
\affiliation{Photonics Initiative, Advanced Science Research Center, City University of New York, New York, New York 10031, USA}

\author{A. Al\`u}
\email{aalu@gc.cuny.edu}
\affiliation{Photonics Initiative, Advanced Science Research Center, City University of New York, New York, New York 10031, USA}
\affiliation{Physics Program, Graduate Center, City University of New York, New York, New York 10016, USA}
\author{J. Yuen-Zhou}
\email{jyuenzhou@ucsd.edu}
\affiliation{Department of Chemistry and Biochemistry, University of California San Diego, La Jolla, California 92093, USA}

\begin{abstract}

We investigate the effect of nonrelativistic motion on the emission dynamics of a dipole emitter moving next to a reflecting interface. Within the formalism of macroscopic QED, we obtain a general equation of motion for the dipole amplitude in terms of the dyadic Green's function, yielding a dynamical extension of the Drexhage effect. At short dipole-surface distances, the dipole can be described as a parametric oscillator featuring time-dependent dampings and Lamb shifts, both arising from the self-induced modulation of the surrounding electromagnetic environment. Importantly, these time-dependent parameters do not always average out, leading to amplification of the dipole amplitude and the radiated intensity when considering certain sinusoidal trajectories with specific modulation amplitudes and frequencies. We derive threshold modulation amplitudes as function of the relative permittivities at the interface. Qualitatively, in the vicinity of certain epsilon-near-zero materials, amplification is possible purely by modulation of the damping. Our findings open up avenues for the dynamic control of light–matter interaction in nanophotonic environments.\\

\vspace{0.5em}
\end{abstract}

\maketitle


\section{Introduction}


Advanced control over the properties of quantum emitters is central to the development of photonic quantum technologies \cite{book_Novotny}. The spontaneous emission of a dipole can be controlled through engineering its electromagnetic environment, a phenomenon first recognized by Purcell \cite{purcell1946,annurev_Bourgeois}. This modification of the Purcell effect is quantified by the local optical density of states (LDOS) \cite{Beutler_RET,Drexhage_influence_surface, Barnes_class_quant_emitters, book_Novotny, Langguth_Drexhage_sound,PhysRevA.110.062812}, which can be obtained from the electromagnetic dyadic Green's function for a given spatial profile of permittivity and permeability \cite{book_Novotny, book_Tai, Guozhong_technq_green,GutierrezReyes2014}. Notably, with the advent of metamaterials, like photonic crystals \cite{BUTT2021107265}, and metasurfaces \cite{Yang_spont_metasurface}, which offer properties absent in natural materials—such as near-zero permittivities, negative refractive indices, and time-dependent permittivities—the ability to control dipole emission has expanded dramatically \cite{11085138,Alhamdan:24,PhysRevB.91.205126,Cui_2024,8107896}. In addition, chemical and materials science developments have made it possible to tailor the emitters themselves by designing energy level separation using different geometries, as in quantum dots, NV centers, and organic molecules \cite{Yang_HighPurity, Duan2003, Johlin2016super, Muravitskaya_enginering_ldos, noda2007spontaneous,Katsumi2025, toninelli2021}. A general theoretical formalism that is suitable for analyzing the radiation of quantum emitters in dissipative and dispersive environments with complex geometry is given by macroscopic quantum electrodynamics (QED), where the presence of macroscopic structures is described by a continuum of collective light-matter (``polaritonic") modes \cite{Feist_macroQED_nanoph,Scheel_MacroscopicQED,BUHMANN_disp_forces,Chuang_TC_macroQED_perspective, Horsley_tutorial_MQED}.\\

Beyond static electromagnetic environments, more recently, advances in Floquet engineering have introduced new ways to tune light-matter interactions via time-periodic modulation of the surrounding medium, thereby unlocking new \textit{temporal} degrees of freedom for controlling radiative processes \cite{Smith_Don_Belyanin,yin2022, galiffi2022, Lyubarov_amp_emission_lasing_ptc, Boltasseva:24, Wang2025}. Time-varying photonic structures—including photonic time crystals—have demonstrated that dynamic control of the LDOS can strongly influence light emission \cite{martinez2016temporal, Lyubarov_amp_emission_lasing_ptc, Asgari:24, park2024spontaneous}, establishing an area of rapidly growing interest in photonics. In this work, we extend the paradigmatic Drexhage problem of a dipole emitter placed in front of a reflecting surface with complex permittivity \cite{Drexhage_influence_surface,annurev_Bourgeois} into the dynamic regime, by considering periodic modulation of the dipole's position. This approach combines the material tunability of metasurfaces with the temporal control of Floquet engineering for the control of radiative processes in quantum emitters.\\

The problem of a dipole changing its position in time represents a distinct class of a time-dependent environment, relevant in both relativistic \cite{Scully_enhancing_acceleration} and nonrelativistic \cite{Lo_quntum_radiation_shaken} regimes, and has been studied in contexts ranging from Fabry-P\'erot cavities \cite{Wilkens_spontaneous_fp} and infinite mirrors \cite{Svidzinsky_excitation_accelerated_mirror} to free space \cite{Wilkens_significance_R_current}. Different motion profiles—including uniform acceleration \cite{Scully_enhancing_acceleration, Svidzinsky_excitation_accelerated_mirror} or oscillatory shaking \cite{Dolan_shaking_photons}—give rise to distinct emission behaviors. We also note that the problem discussed here is closely related to a dynamical spin–boson model with time-dependent coupling to the electromagnetic bath, providing a natural setting to explore how external modulation shapes dissipation in open quantum systems \cite{Peskin_analysis_Langevin-Schr,Peskin_book, PhysRevA.98.052129,Peskin10052012}.\\

\begin{figure}
\includegraphics[scale=0.99]{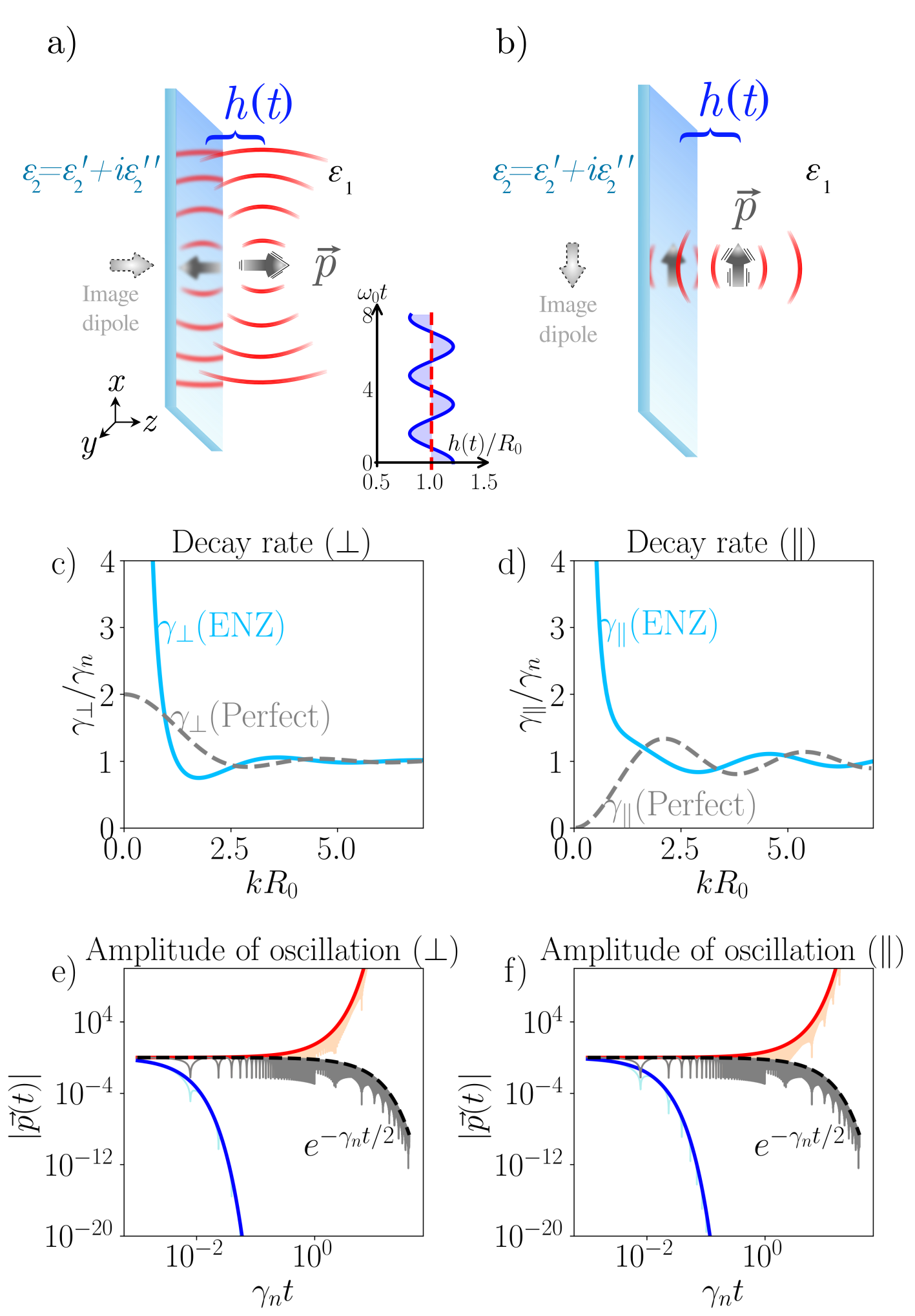}
\caption{  (Color online) a), b) Illustration of dipoles placed in front of infinite mirrors with complex permittivity $\varepsilon_2 = \varepsilon_2' + i \varepsilon_2''$. The dipole moment, depicted by an arrow, is oriented either a) perpendicular ($\perp$) or b) parallel ($\parallel$) to the mirror surface. The distance to the mirror, $h(t)$, is modulated periodically in time, as shown in the inset. The dipoles are embedded in a medium with permittivity $\varepsilon_1$. The corresponding image dipoles and axis orientations are also indicated.
c), d) Static decay rates for perpendicular c) and parallel d) dipole orientations near a mirror, normalized to the homogeneous-medium result $\gamma_n$, as a function of distance to the mirror $R_0$ scaled by the wave number $k$. The gray dashed curves show the result for perfect mirrors ($\varepsilon_2 \to -\infty$), while the solid blue curves show the result for $\varepsilon_2 / \varepsilon_1 = -2^{-1/2} + i 2^{-1/2}$.
e), f) Qualitative time dynamics of modulated dipoles: Parametric amplification is shown by the red curves with an amplitude of modulation of $A= 0.67$. The blue line corresponds to the case of a static position at $(kR_0)^{-1}=12$, while the black curve corresponds to decay in a homogeneous medium with refractive index $n$, $\exp(-\gamma_n t/2)$. Light-colored lines show the qualitative oscillations of the dipole moment (logarithmic scale), while dark-colored lines show the envelope of the absolute value.}
\label{fig_example_sys}
\end{figure}

The case of a moving dipole in front of a perfect mirror was previously analyzed by Belyanin et al.~in Ref.~\cite{Belyanin_coherent_rad_moving_atom}. However, in realistic settings, mirrors are composed of lossy materials, for which the spontaneous decay rate increases sharply as the dipole approaches the surface [see, e.g., blue curves in Figs.~\ref{fig_example_sys} c) and d)], due to strong near-field coupling to evanescent modes at the interface, thereby severely limiting controllability over the emitter’s radiative dynamics. Here, we show that parametric modulation of the dipole's position enables active tuning of the decay rate even in close proximity to such lossy materials.\\


Understanding the effects of dipole motion in emission properties is key to developing novel light-generation mechanisms \cite{Belyanin_coherent_rad_moving_atom}, particularly those arising from its full coupling to the electromagnetic modes described by the vector wave equation. Previous studies of dipole motion have often examined individual modes in isolation, offering fragmented views of the underlying phenomena \cite{Scully_enhancing_acceleration, Dolan_shaking_photons}. Here, within the formalism of macroscopic QED, we present a unified framework that captures the complete coupling of a moving dipole to the entire mode spectrum, enabling a more comprehensive understanding of the resulting dynamical processes. \\

We derive a general equation of motion for the dipole amplitudes with a time-dependent position in an arbitrary structured electromagnetic environment. One of the difficulties is that an electric dipole couples to the electric field in its own frame of reference. In the nonrelativistic regime, we approximate the electric field by the one in the laboratory frame by neglecting terms proportional to the speed of the dipole divided by the speed of light \cite{Wilkens_spurious, book_Jackson_EM}. The resulting equation of motion is given in terms of a general dyadic Green's function that models the presence of a reflecting surface. Within subwavelength separations between the dipole and the mirror, we find a general Markovian equation for the dipole amplitudes, such that retardation effects can be neglected \cite{book_Jackson_EM}. The Markovian equation is given in terms of a dynamical decay rate and a dynamical Lamb shift that depend on the instantaneous position of the dipole. As in the static case, their contributions to the dynamics of the dipole is determined by the permittivity of the mirror. We find that for oscillatory trajectories close to the mirror, as illustrated in Fig.~\ref{fig_example_sys}, the dipole behaves as a parametric oscillator governed by a Mathieu equation, allowing us to identify threshold driving amplitudes of the position that lead to an increase in the amplitude of the dipole. Qualitatively, within the regime of epsilon-near-zero (ENZ) materials \cite{Alu_ENZ-tailoring,arnoldus2019force, Reshef2019, Arnoldus_force_ENZ,Maas_enz_realization}, the dynamical Lamb shift can be suppressed for specific values of the permittivity. In this case, pure modulation of the dynamical decay rate is sufficient to induce dipole amplification, as shown in Figs.~\ref{fig_example_sys} e) and f).\\

The structure of the article is as follows. In Section~\ref{Sec_model_EOM}, we introduce the model of a time-dependent electric dipole oscillator and derive a non-Markovian equation of motion expressed in terms of the dyadic Green’s function for arbitrary electromagnetic environments. We then specialize to the case of an emitter next to a mirror and analyze the Markovian limit by considering dipole trajectories in close proximity to the surface. In Section \ref{Sec_par_osc}, we show that the dipole equation can be mapped onto a parametric oscillator with coefficients determined by the trajectory and the environment. We show that amplification can occur at short distances and derive a general condition for its onset near arbitrary, lossy reflectors characterized by the relative permittivities of the interface materials. \mike{We demonstrate the broad applicability of our results by analyzing amplification across various material regimes, including plasmonic metals, lossy metals, and ENZ materials in Section \ref{Sec_diff_reg}. We discuss possible physical implementations of our proposal in Section \ref{Sec_phys_implement}}. In Section \ref{Sec_fin_rem}, we provide our concluding remarks.\\

\section{Equation of motion of a moving dipole}\label{Sec_model_EOM}

We consider a pointlike emitter, modeled as an electric dipole oscillator $ \hat{\vec{p}}(t) $ of frequency $\omega_0$ and mass $m$. In the Heisenberg picture, the dipole operator is given by \cite{Wylie_quantum_elec_interface2}
\begin{equation}\label{eq_def_dipole_operator}
\hat{\vec{p}}(t) = \vec{n}_{p} q \sqrt{\frac{\hbar }{2 m \omega_0}} [\sigma(t) + \sigma^{\dagger}(t)],
\end{equation}
where we have introduced bosonic annihilation and creation operators that satisfy $[\sigma(t), \sigma^\dagger(t)]=1$, corresponding to a quantum emitter in the weak excitation regime where saturation effects can be neglected. Further, we have introduced the charge $q$ and the unit vector $\vec{n}_{p}$ along the dipole moment $\vec{p}_{q}=\vec{n}_{p}q\sqrt{\hbar/(2 m \omega_0)}$. We assume that the permeability is a real constant $\mu_1$ in the entire domain, while the permittivity is a general complex function of position and frequency, $\varepsilon_1(\vec{x},\omega)$.  The dipole moves along a prescribed trajectory $\vec{h}(t)$ in an arbitrary environment, within the nonrelativistic regime. In addition, in our model the dipole is linearly coupled to the local electric field $\hat{\vec{E}}(t)$. Further details of the Hamiltonian are given in Supplementary Information (SI) \ref{subsec_hamiltonian}. From the Heisenberg equations of motion, the evolution of the expectation value of the dipole operator is given by:
\begin{equation}\label{eq_motion_dipole_v1}
\frac{d^2}{dt^2} \langle \hat{\vec{p}}(t) \rangle + \omega_0^2 \langle \hat{\vec{p}}(t) \rangle = \vec{p}_{q} \frac{2 \omega_0 \vec{p}_{q}}{\hbar} \cdot \langle \hat{\vec{E}}(t) \rangle.
\end{equation}
In the SI, section \ref{subsec_dyadic_green_e}, we show that the expectation value of the electric field can be written in terms of the dyadic Green's function $\overline{\overline{G}}(\vec{r},\vec{x},\omega)$,
\begin{equation}\label{eq_motion_dipole_dyadic}
\begin{split}
&\frac{d^2}{dt^2} \langle \hat{\vec{p}}(t) \rangle + \omega_0^2 \langle \hat{\vec{p}}(t) \rangle \\
&= \frac{1}{2\pi} \int_{- \infty}^{\infty}  d\omega_1   \int_{-\infty}^{\infty} d \tau_1   \;      e^{-i \omega_1  ( t - \tau_1) }     \\
& \Bigg( \frac{2 \omega_0 }{ \hbar } \Bigg) [ \mu_1 \omega_1^2]  \Bigg[   [\vec{p}_{q} \cdot  \overline{\overline{G}}(\vec{h}(t), \vec{h}( \tau_1  ) , \omega_1) \cdot \vec{p}_{q}]  \langle \hat{\vec{p}}(\tau_1) \rangle  \Bigg],
\end{split}
\end{equation}
which allows us to obtain the total electric field that drives the dipole, including the one that is scattered by the surroundings and returns to the dipole at different positions\textemdash an expression that, to the best of our knowledge, has not been reported in the literature. Let us emphasize that, within the macroscopic QED formalism \cite{Feist_macroQED_nanoph,Chuang_TC_macroQED_perspective}, the above formulation is completely general and holds for any non-magnetic environment that can be represented with an electric dyadic Green's function (see SI, section \ref{subsec_dyadic_green_e}). To simplify the notation, we henceforth omit operator hats and expectation-value brackets, by writing $ \langle \hat{\vec{p}}(t) \rangle \rightarrow \vec{p}(t) $ and $\langle \hat{\vec{E}}(t) \rangle \rightarrow \vec{E}(t) $. In contrast to the case of a dipole with fixed position, the Born approximation $ \vec{p}(t)  = \text{Re}[ \vec{p}(0)  e^{-i \omega_0 t}]  $ does not yield a Markovian equation of motion for a moving dipole due to the emergence of a dynamical retarded time \cite{book_Jackson_EM}: the dynamics requires knowledge of previous positions of the dipole.\\

In conventional electrodynamics, non-Markovian behavior typically manifests as retardation effects \cite{book_Jackson_EM}. However, in the present case, the retarded time is not a constant shift in time but instead depends on all the previous positions of the dipole. We shall now illustrate an application of Eq.~\eqref{eq_motion_dipole_dyadic} to the case of a moving dipole in front of an infinite reflecting surface of constant permittivity that fills the half-space $z<0$.\\

In general, the permittivity of the reflecting surface depends on the frequency and on the in-plane wave vector, as required by causality \cite{book_Jackson_EM,Fuchs_nonlocal_metal,Monticone:25}. In the present work, in the weak interaction regime, we consider a permittivity that is approximately constant around the emission frequency of the dipole $\omega_0$, within an interval determined by the maximum of the decay rate or the absolute value of the Lamb shift evaluated at the shortest distance to the mirror, as will be justified below. Deviations from this approximation may modify the behavior of the scattered field in a nontrivial way, such as changing the dependence on the distance, or the appearance of a finite value for the decay rate and Lamb shift. These corrections are expected to be relevant at extremely short distances from the surface \cite{Fuchs_nonlocal_metal}. In what follows we will consider short distances, but still large enough for the constant-permittivity approximation to remain valid.\\

Let us consider trajectories perpendicular to the mirror $\vec{h}(t) = (0,0, h(t))$. In the weak-coupling approximation, where the dipole has small deviations from its free oscillatory behavior given by $\omega_0$, Eq.~\eqref{eq_motion_dipole_dyadic} gives rise to a non-Markovian parametric oscillator:
\begin{equation}\label{eq_general_damping_lamb}
\begin{split}
& \frac{d^2}{dt^2} \vec{p}(t)  + \omega_0^2 \vec{p}(t) \\
& \approx \omega_0 \Delta \upsilon(\vec{h}(t),\vec{h}(t_{\text{prev}})) \vec{p}(t)  - \Gamma(\vec{h}(t),\vec{h}(t_{\text{prev}})) \frac{d}{dt} \vec{p}(t) ,
\end{split}
\end{equation}
where we have introduced a non-Markovian dynamical damping coefficient $\Gamma (\vec{h}(t), \vec{h}(t_{\text{prev}}))$, a dynamical non-Markovian Lamb shift coefficient $\Delta \upsilon(\vec{h}(t), \vec{h}(t_{\text{prev}}))$ and a retarded time $t_{\text{prev}}$. In the following we will specify each of the coefficients and their arguments. The dynamical damping is given by
\begin{equation}\label{eq_aux_decay_rates_imprf-nonMarkovian}
\begin{split}
&\frac{\Gamma (\vec{h}(t), \vec{h}(t_{\text{prev}}))}{\gamma_n} \\
&= \frac{1}{\gamma_n} \frac{6 \pi \varepsilon_1 \omega_0^2 \mu_1}{ k^3} \vec{n}_p \cdot \text{Im}[ \overline{\overline{G}}(\vec{h}(t),\vec{h}(t_{\text{prev}}),\omega_0)  ] \cdot \vec{n}_p\\
&= 1 + \frac{6 \pi \varepsilon_1 \omega_0^2 \mu_1}{ k^3} \vec{n}_p \cdot \text{Im}[ \overline{\overline{G}}_{s}(\vec{h}(t),\vec{h}(t_{\text{prev}}),\omega_0)  ] \cdot \vec{n}_p.
\end{split}
\end{equation}
Here, $k$ represents the wave number defined by $\omega_0 n = c k$ in terms of the refractive index $n = \sqrt{\varepsilon_1 \mu_1} / \sqrt{\varepsilon_0 \mu_0}$, with $\varepsilon_1$ and $\mu_1$ the permittivity and permeability that surround the dipole, while $\varepsilon_0$ and $\mu_0$ are the corresponding vacuum constants. In the last equality we have separated the dyadic Green's function as
\begin{equation}
\overline{\overline{G}} (\vec{x},\vec{y}, \omega) = \overline{\overline{G}}_0 (\vec{x},\vec{y}, \omega) + \overline{\overline{G}}_s (\vec{x},\vec{y}, \omega), 
\end{equation} 
with the first term corresponding to the homogeneous-space dyadic Green's function, and the last term spans the reflected electric field that was scattered by the surroundings \cite{book_Novotny}. In Eq.~\eqref{eq_aux_decay_rates_imprf-nonMarkovian} we have also introduced the decay rate in homogeneous space:
\begin{equation}
\gamma_n = \frac{6 \pi \varepsilon_1 \omega_0^2 \mu_1}{ k^3} \vec{n}_p \cdot \text{Im}[ \overline{\overline{G}}_{0}(\vec{x},\vec{x},\omega_0)  ] \cdot \vec{n}_p,
\end{equation}
where $\vec{x}$ is an arbitrary position. The dynamical non-Markovian Lamb shift coefficient is
\begin{equation}\label{eq_aux_Lamb_shift_imprf-nonMarkovian}
\begin{split}
&\frac{\Delta \upsilon(\vec{h}(t), \vec{h}(t_{\text{prev}}))}{\gamma_n} \\
&= \frac{6 \pi \varepsilon_1 \omega_0^2 \mu_1 }{k^3}    \vec{n}_p \cdot  \text{Re}[ \overline{\overline{G}}_{s}(\vec{h}(t), \vec{h}(t_{\text{prev}}),\omega_0)  ]\cdot \vec{n}_p.
\end{split}
\end{equation}
As in the case of a dipole with fixed position \cite{book_Novotny}, the dynamical Lamb shift $\Delta \upsilon(\vec{h}(t), \vec{h}(t_{\text{prev}}))$ is given in terms of $\overline{\overline{G}}_{s}(\vec{h}(t), \vec{h}(t_{\text{prev}}),\omega_0)$ and there is no homogeneous space contribution $\overline{\overline{G}}_0(\vec{h}(t), \vec{h}(t_{\text{prev}}),\omega_0)$ because the speed of the dipole is negligible compared to the speed of light. That is, the dipole does not move fast enough to re-encounter the light it emits, only reflected radiation can return to the dipole. We have also introduced the retarded time $t_{\text{prev}}$, at which light was emitted in the past, that is determined by solving the equation
\begin{equation}\label{eq_t_prev}
t_{\text{prev}} = t - n |h (t) + h(t_{\text{prev}})|/c.
\end{equation}
This equation is valid for any reflecting surface since it arises from the general structure of the dyadic Green's function. Physically, it describes the propagation of light, in a medium of constant refractive index $n$, traveling from a previous time $t_{\text{prev}}$ to a present time $t$ upon reflection from the mirror, see SI \ref{subsec_ref_elec_field_perf}  for the graphical solution of Eq.~\eqref{eq_t_prev}. In a more general environment, there could be more retarded times, depending on the number and shape of objects that can scatter light. For example, if we intersect two infinite mirrors, we would expect several retarded times due to reflection between mirrors.\\

As an illustrative example, let us consider the case of a perfect mirror and a dipole perpendicular ($\perp$) to the mirror, see Fig.\ref{fig_example_sys} a). The contribution from the reflected electric field $\vec{E}_{s \perp}(t)$ generated by $\overline{\overline{G}}_s (\vec{x},\vec{y}, \omega)$ is given by
\begin{equation}\label{eq_electric_field_moving_dip_perp_main-text}
\begin{split}
\vec{p}_{q} \frac{2 \omega_0 \vec{p}_{q}}{\hbar} \cdot  \vec{E}_{s \perp}(t)  = & \frac{q^2}{m} \frac{1}{4 \pi \varepsilon_1} \Bigg[ \frac{2  \vec{p}(t_{\text{prev}})  }{[h(t)+h(t_{\text{prev}})]^3} \\
&+ \frac{n}{c}\frac{2 \frac{d}{dt}  \vec{p}(t_{\text{prev}})  }{[h(t)+h(t_{\text{prev}})]^2}  \Bigg],
\end{split}
\end{equation}
where $ t_{\text{prev}} $ is given in Eq.~\eqref{eq_t_prev}. This expression can be obtained from the dyadic Green's function or also using the image method \cite{book_Novotny}. Although the structure of the scattered electric field is the same as in the fixed-position case \cite{book_Novotny}, we see a non-Markovian behavior, since the field was generated by an image dipole at the previous time $t_{\text{prev}}$ at a time-dependent distance $h(t) + h(t_{\text{prev}})$. When we use the Born approximation $\vec{p}(t) = \vec{p}(0) \text{exp}[-  i \omega_0 t]$ we can split the expression into real and imaginary terms to obtain the scattered-field contribution of the non-Markovian coefficients given in Eqs.~\eqref{eq_aux_decay_rates_imprf-nonMarkovian} and \eqref{eq_aux_Lamb_shift_imprf-nonMarkovian}. Indeed, in the fixed-position case $h(t) = h$ we can recover the well-known contributions to the Lamb shift and decay rate.\\

We can obtain a Markovian dynamics by considering an instantaneous feedback approximation, which is valid if the dipole moves slowly compared to the time required for the emitted light to return back to the dipole. Using a coarse-graining method with a time step $\Delta t$ larger than the travel time of the light emitted from the dipole and reflected by the mirror $\Delta t \gtrsim \max \{2 n h(t) /c\} $ such that $h(t) \approx h(t_{\text{prev}})$, we can obtain a Markovian equation with time-dependent coefficients,
\begin{equation}\label{eq_parametric_osc_diff}
\frac{d^2}{dt^2} \vec{p}(t)  + \omega_0^2 \vec{p}(t)  \approx \omega_0 \Delta \nu(\vec{h}(t)) \vec{p}(t)  - \gamma(\vec{h}(t)) \frac{d}{dt} \vec{p}(t) ,
\end{equation}
where the dynamical damping is
\begin{equation}\label{eq_aux_decay_rates_imprf}
\gamma (\vec{h}(t)) = \Gamma (\vec{h}(t), \vec{h}(t))
\end{equation}
and the dynamical Lamb shift is
\begin{equation}\label{eq_aux_Lamb_shift_imprf}
\Delta \nu (\vec{h}(t)) =  \Delta \upsilon(\vec{h}(t), \vec{h}(t )).
\end{equation}
A key aspect of the coefficients given in Eqs.~\eqref{eq_aux_decay_rates_imprf} and \eqref{eq_aux_Lamb_shift_imprf} is that they correspond to parameterizations of the well-known static coefficients by considering the instantaneous position of the dipole $\vec{h}(t)$ \cite{book_Novotny, Chance_lifetime_emitting_molecule}. Hence, it is straightforward to recover the non-modulated case when we consider a static position in Eqs.~\eqref{eq_aux_decay_rates_imprf-nonMarkovian} and \eqref{eq_aux_Lamb_shift_imprf-nonMarkovian} and in Eqs.~\eqref{eq_aux_decay_rates_imprf} and \eqref{eq_aux_Lamb_shift_imprf}. Also, we can recover the established results of the non-modulated case when we consider a static position in Eq.~\eqref{eq_motion_dipole_dyadic} and use the Born approximation, as shown in section \ref{subsec_eq_motion_static_field} of the SI. In the rest of this work, we will focus on the dynamics described by Eq.~\eqref{eq_parametric_osc_diff}, which is a Markovian equation valid for motion of the dipole at short distances from the mirror.\\

From the solutions of Eq.~\eqref{eq_parametric_osc_diff}, we can calculate the lifetime $\tau$ of the dipole operator, which is defined when the dipole exhibits exponential decay in time,
\begin{equation}\label{eq_invs_lifetime}
\vec{p}(t)  \propto \vec{p}_0(t)  e^{- t /\tau},
\end{equation}
where $\vec{p}_0(t)$ is an oscillatory function. The relation between the inverse lifetime and the decay rate obtained from Fermi's golden rule $\gamma_{F}$ is given by $\tau = 2 / \gamma_F$. However, as we will show later, in the cases where we amplify the dipole moment, this relation has to be reinterpreted due to the sign flip of $\tau$ \cite{Lyubarov_amp_emission_lasing_ptc, Puneet_inv-design, Juanjuan_classical-purcell}.\\

\section{Parametric oscillator}\label{Sec_par_osc}
We will consider separately the cases of a perpendicular dipole $(\perp)$ and a parallel dipole $(\parallel)$, \mike{as sketched in Figs.~\ref{fig_example_sys} a), b)}. \textcolor{black}{For a weakly interacting dipole, we can characterize an imperfect mirror by a complex permittivity $\varepsilon_2 = \varepsilon_2' + i \varepsilon_2''$ that is approximately constant around the emission frequency $\omega_0$.} The coefficient functions $\gamma (\vec{h}(t))$ and $\Delta \nu (\vec{h}(t))$, given in Eqs.~\eqref{eq_aux_decay_rates_imprf} and \eqref{eq_aux_Lamb_shift_imprf}, correspond to the parametrization of the static near-field approximation. That is, in contrast to the case of a perfect mirror, we cannot calculate the reflected electric field explicitly, as shown in Eq.~\eqref{eq_electric_field_moving_dip_perp_main-text}. However, from our analysis of the perfect mirror, we can identify that the key properties that allow us to obtain the instantaneous dynamical Lamb shift and decay rate are the same: The closeness to the mirror (to obtain an instantaneous feedback) and the weak interaction regime (to obtain the dynamical Lamb shift and decay rate). \textcolor{black}{Then, at short distances to the mirror the reflected electric field corresponds to the near-field emitted by the dipole.} This is shown in the SI \ref{subsec_elec_field_imperf}. Thus, in the short distance limit, we obtain the differential equation of a parametric oscillator:
\begin{equation}\label{eq_diff_eq_dipole_kb}
\begin{split}
&\frac{d^2}{dt^2} \vec{p}(t)  + \omega_0^2 \vec{p}(t) \\
& \approx  \frac{K_{\perp, \parallel}}{(2 h(t))^3} \vec{p}(t)  - \Bigg( \frac{b_{\perp, \parallel}}{(2 h(t))^3} + \gamma_{t \perp, \parallel} \Bigg) \frac{d}{dt} \vec{p}(t),
\end{split}
\end{equation}
where, as stated before, we consider motion perpendicular to the mirror given by $h(t)$, the time-dependent distance to the mirror. For small enough distances, we can make the following approximations for the coefficients, see SI \ref{subsec_elec_field_imperf}, for a dipole moment perpendicular to the surface:
\begin{equation}\label{eq_K_perp}
K_{\perp} = \frac{q^2}{m} \text{Re} [  R(\varepsilon_2/\varepsilon_1)   ]  \frac{2}{(4 \pi) \varepsilon_1}
\end{equation}
and
\begin{equation}\label{eq_b_perp}
b_{\perp} = \frac{q^2}{m \omega_0} \text{Im}  [  R(\varepsilon_2/\varepsilon_1) ]  \frac{2}{(4\pi)\varepsilon_1},
\end{equation}
where we have defined the function
\begin{equation}\label{eq_R_varepsilon}
R(\varepsilon_2/\varepsilon_1) = \frac{\varepsilon_2 - \varepsilon_1 }{\varepsilon_2 + \varepsilon_1 }, 
\end{equation}
which corresponds to a Fresnel-like coefficient at normal incidence \cite{book_Novotny}, while for a dipole parallel to the surface we have
\begin{equation}\label{eq_Kb_parall}
K_{\parallel} = \frac{K_{\perp}}{2}, \quad b_{\parallel} = \frac{b_{\perp}}{2}.
\end{equation}
The damping constants $\gamma_{t \perp, \parallel}$ also depend on the permittivity of the mirror. When $\text{Im}[\varepsilon_2] \neq 0$, we should use $\gamma_{t \perp, \parallel} = \gamma_n$ to account for the damping in free space \cite{book_Novotny}. For the case $\text{Im}[\varepsilon_2] = 0$, we obtain $b_{\perp, \parallel}=0$, and the effective damping due to the reflected electric field (including the far- to near-field contributions) becomes a constant that has to be added to the free space constant damping, giving $\gamma_{t \perp}/\gamma_n \approx 2$ or $\gamma_{t \parallel}/\gamma_n \approx 0$ for the perpendicular and parallel cases, respectively [see gray dashed curves in Figs.~\ref{fig_example_sys} c), d)]. These results correspond to constructive and destructive dipole emission and can be easily understood from the method of images~\cite{Drexhage_influence_surface,Chance_lifetime_emitting_molecule}. Numerical values of the constants $\gamma_{t \perp, \parallel}$ can be calculated for more accuracy, as shown in Figs.~\ref{fig_coef_ideal_mirror} a) and b) \cite{book_Novotny}.\\

We can separate the effect of the damping by means of a canonical transformation, defining a function $ \vec{r}(t) $ as
\begin{equation}\label{eq_transf_exp_increase}
\vec{p}(t)  = \text{exp} \Bigg[ - \int_0^{t} d s\; \frac{1}{2} \Bigg( \frac{b_{\perp, \parallel}}{(2 h(s))^3} + \gamma_{t \perp, \parallel} \Bigg) \Bigg]  \vec{r}(t)  .
\end{equation}
Substituting Eq.~\eqref{eq_transf_exp_increase} into the equation of motion \eqref{eq_diff_eq_dipole_kb}, we find that $\vec{r}(t)$ obeys a differential equation of the form:
\begin{equation}\label{eq_parametric_osc_r}
\frac{d^2}{dt^2} \vec{r}(t)  + \lambda^2(t) \vec{r}(t) = 0,
\end{equation}
where $\lambda (t)$ depends on the constants $K_{\perp, \parallel}$, $b_{\perp, \parallel}$, $\gamma_{t \perp, \parallel}$ and the trajectory $h(t)$. Then, for the cases of a dipole perpendicular or parallel to the surface, we can find conditions on the trajectory $h(t)$ that allow us to obtain a Mathieu equation in Eq.~\eqref{eq_parametric_osc_r} by imposing $\lambda^2(t) = \omega_0^2 [1 + 2 \alpha \text{cos}(\Omega t)]$, with $\alpha$ a parameter to be determined and $\Omega$ the frequency of modulation. Our interest in the Mathieu equation stems from the well known parametric resonances that can increase the amplitude of the dipole oscillations. We consider a modulation of the position given by
\begin{equation}\label{eq_modulation_position}
h(t) = R_0 [1 + \eta(t) ],
\end{equation}
where $|\eta(t)| < 1$ and $R_0$ is an average distance close to the mirror. Substituting Eq.~\eqref{eq_modulation_position} into 
Eqs.~\eqref{eq_transf_exp_increase} and \eqref{eq_parametric_osc_r}, 
and performing a series expansion in $\eta(t)$ while neglecting the quadratic terms $\eta^2(t)$, we find that $\eta(t)$ must satisfy an inhomogeneous ordinary differential equation in order to obtain a Mathieu equation, see SI \ref{subsec_motion_infront_mirror}. When $\text{Im}[\varepsilon_2] \neq 0$, corresponding to the case of an imperfect mirror, the solution of the inhomogeneous differential equation is approximately given by 
\begin{equation}\label{eq_eta_func_t}
\begin{split}
\eta (t) =& \eta_0 e^{-c_{K} \omega_0 t /m_b} + A(\alpha_{\perp,\parallel} ) \text{cos}(\Omega t - \delta),
\end{split}
\end{equation}
where we have defined the amplitude of modulation:
\begin{equation}\label{eq_amplitude_modulation}
A(\alpha_{\perp,\parallel} ) = \frac{2 \alpha_{\perp,\parallel} }{ \sqrt{c_{K}^2 +  (m_b \Omega/\omega_0)^2} },
\end{equation}
as well as the phase shift
\begin{equation}
\delta = \text{tan}^{-1} \Bigg( \frac{\Omega m_b}{\omega_0 c_K } \Bigg).
\end{equation}
Also, we have defined constants for each orientation of the dipole,
\begin{equation}
c_{K} = \frac{3   K_{\perp, \parallel} }{8 \omega_0^2 R_0^3}, \qquad
m_b = \frac{3 b_{\perp, \parallel} }{16 \omega_0 R_0^3}.
\end{equation}
Eq.~\eqref{eq_eta_func_t} determines the trajectory to obtain a Mathieu equation. The first term represents a small deviation from the average distance $\eta_0$, which decays exponentially, while the second term introduces a time-periodic modulation at frequency $\Omega$. \textcolor{black}{Let us emphasize that in the weak interaction regime, dipole emission will be close to its frequency $\omega_0$, while the modulation frequency $\Omega$ affects the values of both dynamical Lamb shift and dynamical damping. Then, the variation of the frequency of emission should not exceed the range in which the permittivity can be considered a constant. Since for lossy materials the Lamb shift and damping increase when we reduce the distance to the mirror, as in Fig.~\ref{fig_example_sys} c) or d), it is enough to use the values at the closest distance to determine the interval around $\omega_0$ where the permittivity should be approximately constant.}\\

Of particular interest is the regime in which the amplitude of the dipole grows exponentially in time. Parametric amplification can occur at modulation frequencies of $\Omega = 2 \omega_0 / M$ for $M=1,2,3,\hdots$. However, for simplicity, we will consider the value of $\Omega = 2 \omega_0$ to analyze parametric amplification. For this frequency of modulation, the dipole amplitude is modulated by \cite{Kovacic_Mathieu_eq}
\begin{equation}\label{eq_dipole_amplitude}
\vec{p}(t)  \propto \text{exp} \Bigg[ \frac{1}{2} \Bigg( \alpha \omega_0 - \frac{b_{ \perp, \parallel}}{8 R_0^3} - \gamma_{ t \perp, \parallel}  \Bigg) t   \Bigg],
\end{equation}
where we are neglecting oscillatory terms that depend on $\eta(t)$. We will show amplification is possible for a suitable value of $\alpha$. For a slightly detuned modulation frequency $\Omega = 2\omega_0 + \epsilon$, the term $\alpha \omega_0$, which governs the exponential growth in Eq.~\eqref{eq_dipole_amplitude}, is replaced by $\sqrt{(\alpha \omega_0)^2 - \epsilon^2}$.\\

\begin{figure}
\includegraphics[scale=0.4]{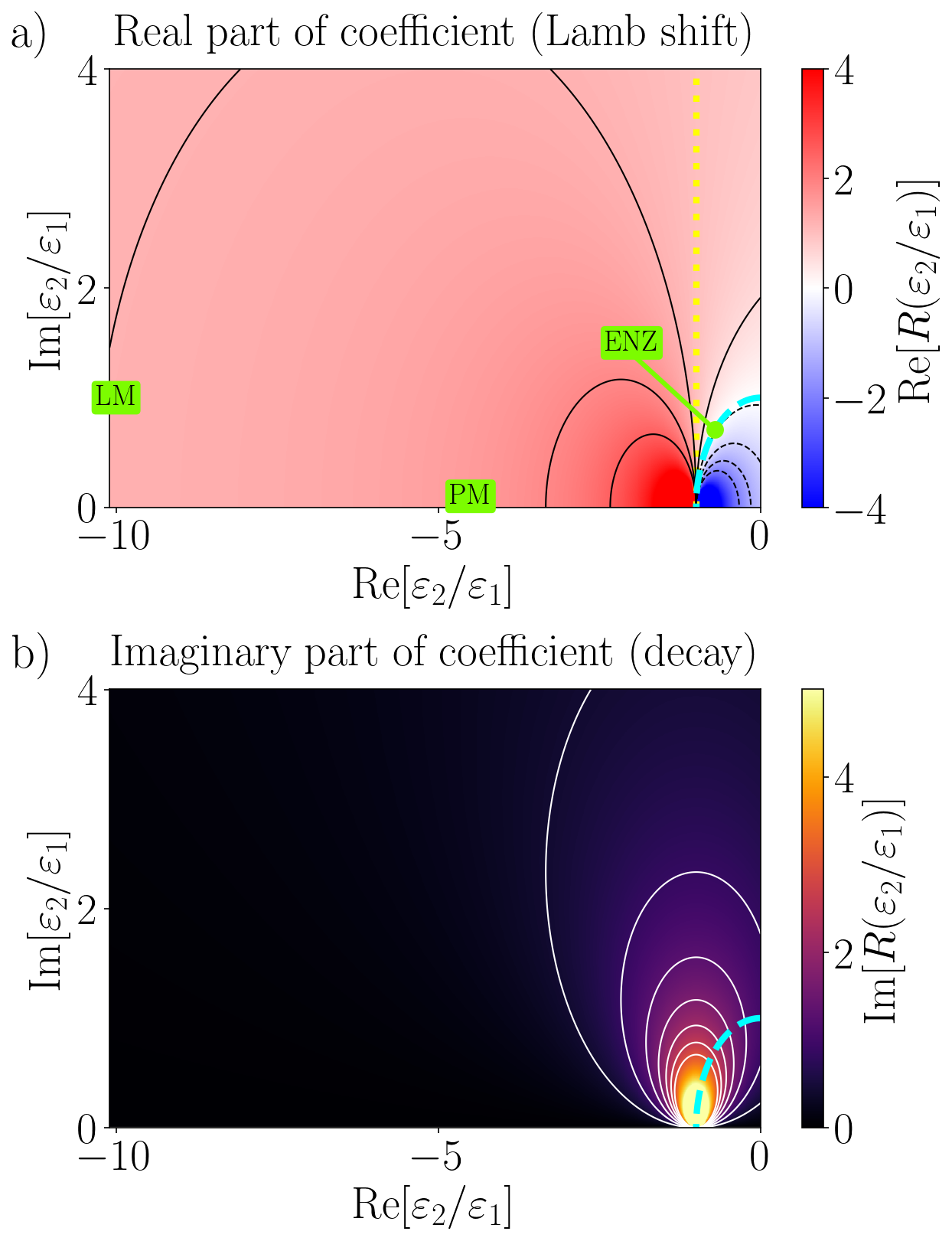}
\caption{(Color online) a), b) \mike{Density plots of the quotient $R(\varepsilon_2 / \varepsilon_1)$ that determines the dynamical frequency shift [real part, a)] and damping [imaginary part, b)] coefficients, according to Eq.~\eqref{eq_R_varepsilon}, plotted as a function of the real and imaginary parts of $\varepsilon_2 / \varepsilon_1$. The dotted vertical yellow line in a) shows the ENZ regime $-1 \leq \text{Re}[\varepsilon_2 / \varepsilon_1] <0$. The dashed cyan curve indicates the permittivities where the dynamical Lamb shift vanishes. In panel a) exemplary points corresponding to different material regimes discussed in the text [lossy metal (LM), plasmonic metal (PM), ENZ material] are labeled for reference. } }
\label{fig_real_imag_R}
\end{figure}

At frequency $\Omega = 2 \omega_0$, we can obtain an increase in the amplitude of the dipole, given approximately in Eq.~\eqref{eq_dipole_amplitude}, if
\begin{equation}\label{eq_threshold_alpha}
\frac{b_{\perp, \parallel}}{8 \omega_0 R_0^3} + \frac{\gamma_{t \perp, \parallel}}{\omega_0} < \alpha_{\perp, \parallel}.
\end{equation}
\mike{However, the modulation amplitude $A(\alpha_{\perp,\parallel})$ must remain below unity to respect the physical boundary imposed by the presence of the mirror.} From the oscillatory term in Eq.~\eqref{eq_eta_func_t}, using $\eta_0=0$, we find an upper boundary for $\alpha_{\perp, \parallel}$ given by
\begin{equation}\label{eq_upper_alpha}
A(\alpha_{\perp,\parallel}) < 1,
\end{equation}
which yields an upper value for any frequency $\Omega$. The approximations, however, are more accurate in the small amplitude regime $A(\alpha_{\perp, \parallel}) \ll 1$. Combining Eqs.~\eqref{eq_amplitude_modulation}, \eqref{eq_threshold_alpha} and \eqref{eq_upper_alpha}, a value for $\alpha_{\perp, \parallel}$ that exhibits amplification can be found when
\begin{equation}\label{eq_finding_amplification}
-\frac{5}{9}m_{b}^2 + \Bigg(\frac{\gamma_{t \perp, \parallel}}{\omega_0} \Bigg)^2 + \frac{4}{3}m_b \frac{\gamma_{t \perp, \parallel}}{\omega_0} < \frac{c_{K}^2}{4}.
\end{equation}
If we consider a short distance regime $(kR_0) \ll 1$, such that the constant damping $\gamma_{t \perp, \parallel}$ in Eqs.~\eqref{eq_threshold_alpha} and \eqref{eq_finding_amplification} is negligible, we find that Eq.~\eqref{eq_finding_amplification} is satisfied. The short distance regime amplification is possible for a wide range of values of permittivity ratios $\varepsilon_2 / \varepsilon_1$, which determine the value of $R(\varepsilon_2 / \varepsilon_1)$ and therefore of $K_{\perp, \parallel}$ and $b_{\perp, \parallel}$, see Eq.~\eqref{eq_R_varepsilon}. In Figs.~\ref{fig_real_imag_R} a) and b), we analyze the behavior of $R(\varepsilon_2 / \varepsilon_1)$. Its real part is associated with the dynamical Lamb shift and its imaginary part is related to the dynamical damping, see Eqs.~\eqref{eq_diff_eq_dipole_kb}, \eqref{eq_K_perp} and \eqref{eq_b_perp}. In particular, we find a region where the dynamical Lamb shift becomes negligible, $\text{Re}[R(\varepsilon_2/\varepsilon_1)]=0$, shown with a dashed (cyan) unit circle in Fig.~\ref{fig_real_imag_R} a), which lies within the ENZ regime \mike{where the real part of the permittivity nearly vanishes, illustrated by the vertical dotted yellow line} \cite{Arnoldus_force_ENZ, Maas_enz_realization}. However, \mike{despite the absence of frequency modulation}, amplification is still possible, according to Eqs.~\eqref{eq_threshold_alpha} and \eqref{eq_upper_alpha}. That is, a rapid change in the dynamical damping coefficient allows us to obtain amplification. To the best of our knowledge, amplification by modulation of friction is a property that has been scarcely explored experimentally  \cite{Perego2018,Krechetnikov_dissipation-induced}, although it has been analyzed theoretically \cite{Scapolan_energy_harvesting} and it is well known that one can map the harmonic oscillator with dynamical friction to a Mathieu equation with an integrating factor \cite{adams2010calculus}.\\

\section{Amplification across different material regimes}
\label{Sec_diff_reg}

\mike{Having derived general conditions and expressions under which dipole amplification can occur via parametric modulation of its position, we now analyze in more detail the minimum amplitude of modulation required to obtain amplification of the electric dipole oscillator, depending on the relative permittivities $\varepsilon_2/\varepsilon_1$. To this end, we analyze a few representative material regimes for the reflecting surface, namely perfect mirrors, lossy metals, plasmonic metals, as well as ENZ materials and analyze how their specific optical properties shape the amplification behavior.}

\subsection{Perfect mirror}

Let us first recover the limiting case of a perfect mirror such as discussed by Belyanin et al.~in Ref.~\cite{Belyanin_coherent_rad_moving_atom}, where $\text{Re}[\varepsilon_2] \rightarrow - \infty$ and $\text{Im}[\varepsilon_2] =0$ such that $b_{\perp,\parallel}=0$ with $\gamma_{t\perp} = 2 \gamma_n$ and $\gamma_{t \parallel} = 0$. We can use Eqs.~\eqref{eq_parametric_osc_r} and \eqref{eq_modulation_position} to identify a trajectory given by $\eta(t) = A(\alpha_{\perp, \parallel}) \, \text{cos}(\Omega t)$, in order to obtain a Mathieu equation that agrees with Ref.~\cite{Belyanin_coherent_rad_moving_atom}. Then, we find 
\begin{equation}\label{eq_alpha_perf_mirr}
A(\alpha_{\perp, \parallel} ) = \frac{16 \omega_0^2 R_0^3 }{3  K_{\perp, \parallel}} \alpha_{\perp, \parallel}.
\end{equation}
From the condition of amplification for a perpendicular dipole, given in Eq.~\eqref{eq_threshold_alpha}, we find the threshold amplitude of modulation reported in \cite{Belyanin_coherent_rad_moving_atom}:
\begin{equation}\label{eq_a_perp_position}
A_{\text{th}\perp} = \Bigg(\frac{32}{9}\Bigg) \frac{n^3 R_0^3 \omega_0^3}{c^3}= \frac{32}{9} (k R_0)^3.
\end{equation}
The threshold amplitude increases with the distance to the mirror, and also increases with the refractive index of the medium that surrounds the dipole. On the other hand, for a parallel dipole, the amplification is given by \cite{Belyanin_coherent_rad_moving_atom}
\begin{equation}\label{eq_a_parallel_position}%
A_{\text{th} \parallel} = 0.
\end{equation}
This last condition comes from the fact that close to a perfect mirror the damping is negligible, as shown by the dashed line in Fig.~\ref{fig_example_sys} d).\\

\subsection{Lossy metal}

\begin{figure}
\includegraphics[scale=0.4]{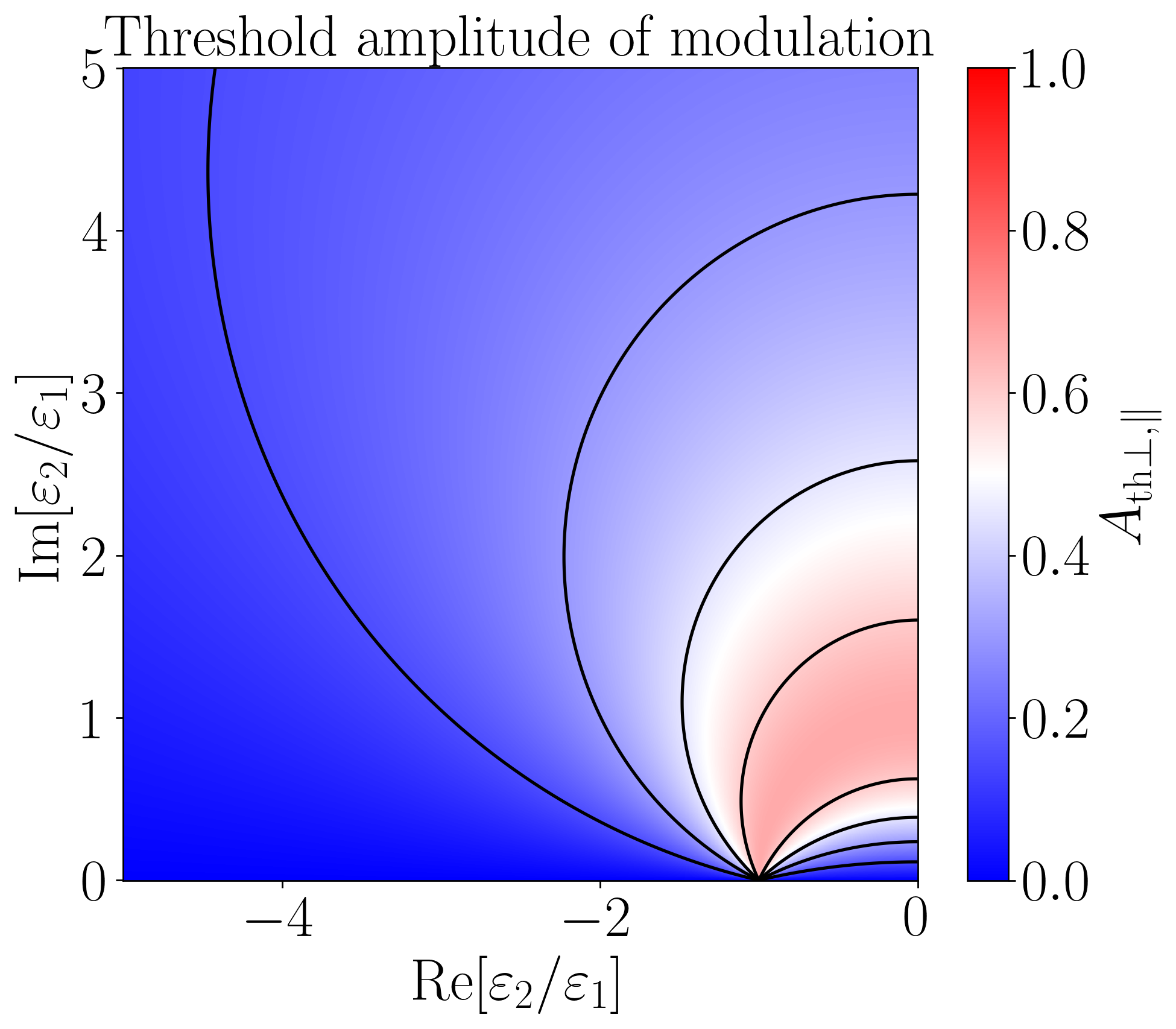}
\caption{(Color online) Threshold amplitude of modulation $A_{\mathrm{th} \perp, \parallel}$ required to achieve amplification at a distance of $(k R_0)^{-1}=12$, as given by Eq.~\eqref{eq_thresh_amp_LM} for a lossy metal. This amplitude defines the trajectory of the dipole given in Eqs.~\eqref{eq_modulation_position} and \eqref{eq_eta_func_t}.}
\label{fig_thresh_ampl_LM}
\end{figure}

The threshold amplitude of modulation for the case of a lossy mirror can be analyzed in general, provided we restrict ourselves to the short distance regime $k R_0 \ll 1$ while considering $\text{Re}[\varepsilon_2] < 0$ and $\text{Im}[\varepsilon_2] \neq 0$. In this case, the constant decay corresponds to $\gamma_{t\perp, \parallel} = \gamma_n$, in contrast with the perfect mirror case. We can obtain the threshold amplitude of modulation from Eqs.~\eqref{eq_amplitude_modulation} and \eqref{eq_threshold_alpha}:
\begin{equation}\label{eq_thresh_amp_LM}
A_{\text{th} \perp, \parallel} = \frac{\frac{2}{3} \zeta_{\perp,\parallel} \text{Im}[R(\varepsilon_2/\varepsilon_1)] + 1 }{ \zeta_{\perp,\parallel} \sqrt{\text{Re}[R(\varepsilon_2/\varepsilon_1)]^2 + \text{Im}[R(\varepsilon_2/\varepsilon_1)]^2 }}.
\end{equation}
where $\zeta_{\perp} = (9/16)/(kR_0)^3$ and $\zeta_{\parallel}= \zeta_{\perp}/2$. Notice that in the short distance regime $A_{\text{th}\perp} \approx A_{\text{th}\parallel}$ since $\zeta_{\perp,\parallel} \gg 1$. Fig.~\ref{fig_thresh_ampl_LM} shows the threshold values obtained for $(kR_0)^{-1} = 12$. The values do not change significantly for shorter distances, as shown in Fig.~\ref{fig_panels_thresh_dis} of the SI. Hence, we can identify regions where a significant modulation is required, such as the ENZ regime, which we will discuss further below. The numerator in Eq.~\eqref{eq_thresh_amp_LM} represents the static damping terms due to the mirror and due to the material with permittivity $\varepsilon_1$. The denominator corresponds to the contributions of the dynamical Lamb shift and dynamical damping, which lower the threshold amplitude of modulation.\\

To illustrate this effect with an example, consider emission at $614\,\text{nm}$ in free space ($\varepsilon_1 = \varepsilon_0$), as in Drexhage’s original experiment carried out with $\mathrm{Eu}^{3+}$ ions \cite{Drexhage_influence_surface}. For a silver interface with $\varepsilon_2 / \varepsilon_1 = -10 + i$ \cite{WINSEMIUS1975529}, the threshold amplitude of modulation is found to be $A_{\text{th} \perp, \parallel} \approx 0.015$. 

Although ENZ materials also fall into this same classification, we defer their discussion to Sec.~\ref{sec:enz} below, to highlight the interesting properties that arise in this
regime, in particular, the possibility of amplification via modulation of
friction.\\

\begin{figure}
\includegraphics[scale=0.3]{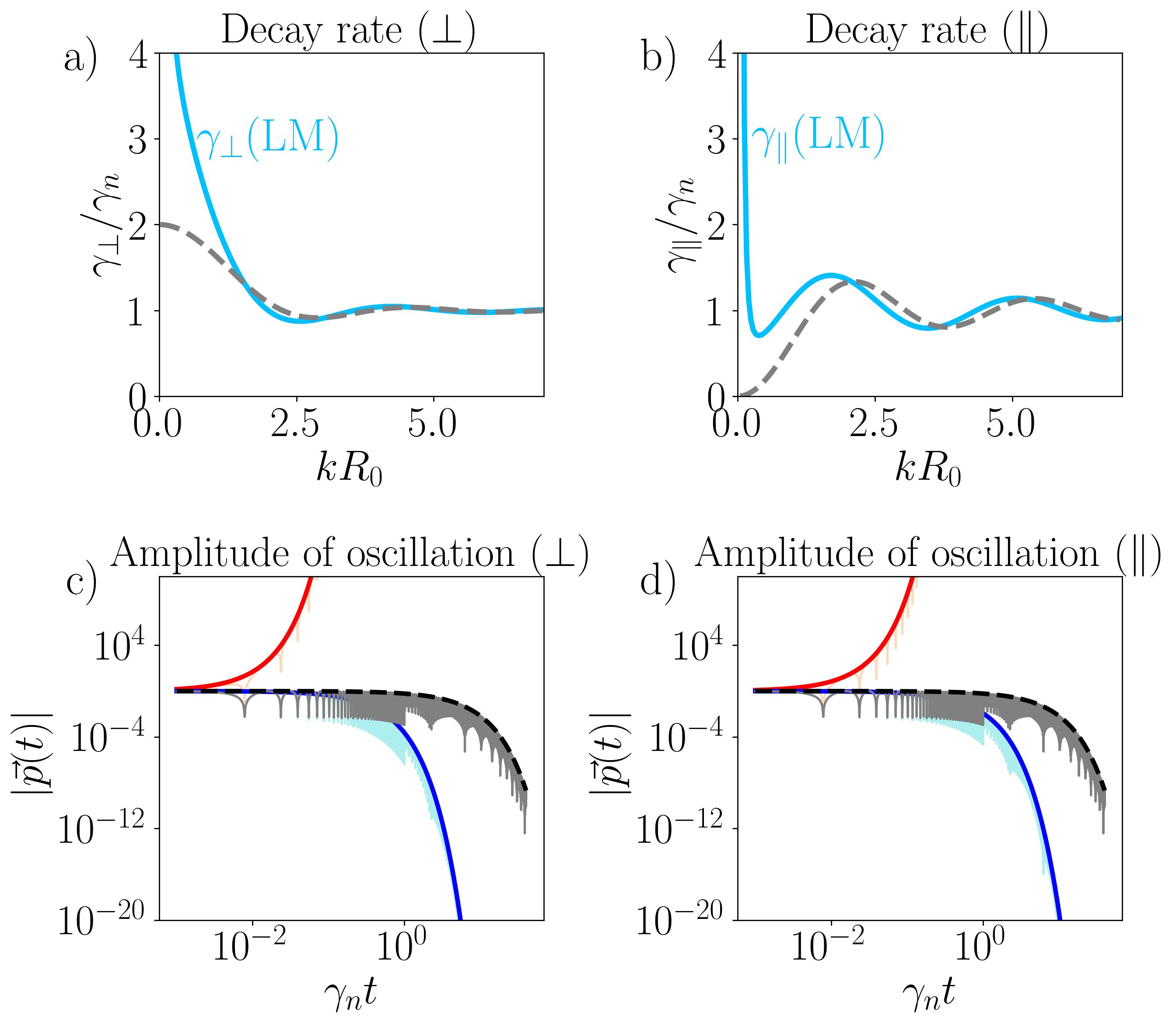}
\caption{(Color online) a), b) Decay rates for a lossy metal (LM) for the a) perpendicular and b) parallel orientations with $\varepsilon_2 / \varepsilon_1 = -10 + i$, shown with solid blue lines. For reference, we include the decays near a perfect mirror, shown with dashed lines. c), d) Qualitative behavior of the absolute value of the oscillations of a dipole on a logarithmic scale. The amplification case corresponds to a modulation amplitude of $A=0.67$ in Eq.~\eqref{eq_eta_func_t}. We chose a distance of $(k R_0)^{-1} = 12$. The blue line corresponds to the case of a static position, while the black curve corresponds to decay in the homogeneous medium.}
\label{fig_decay_osc_LM}
\end{figure}

\subsection{Plasmonic metal}

\begin{figure}
\includegraphics[scale=0.3]{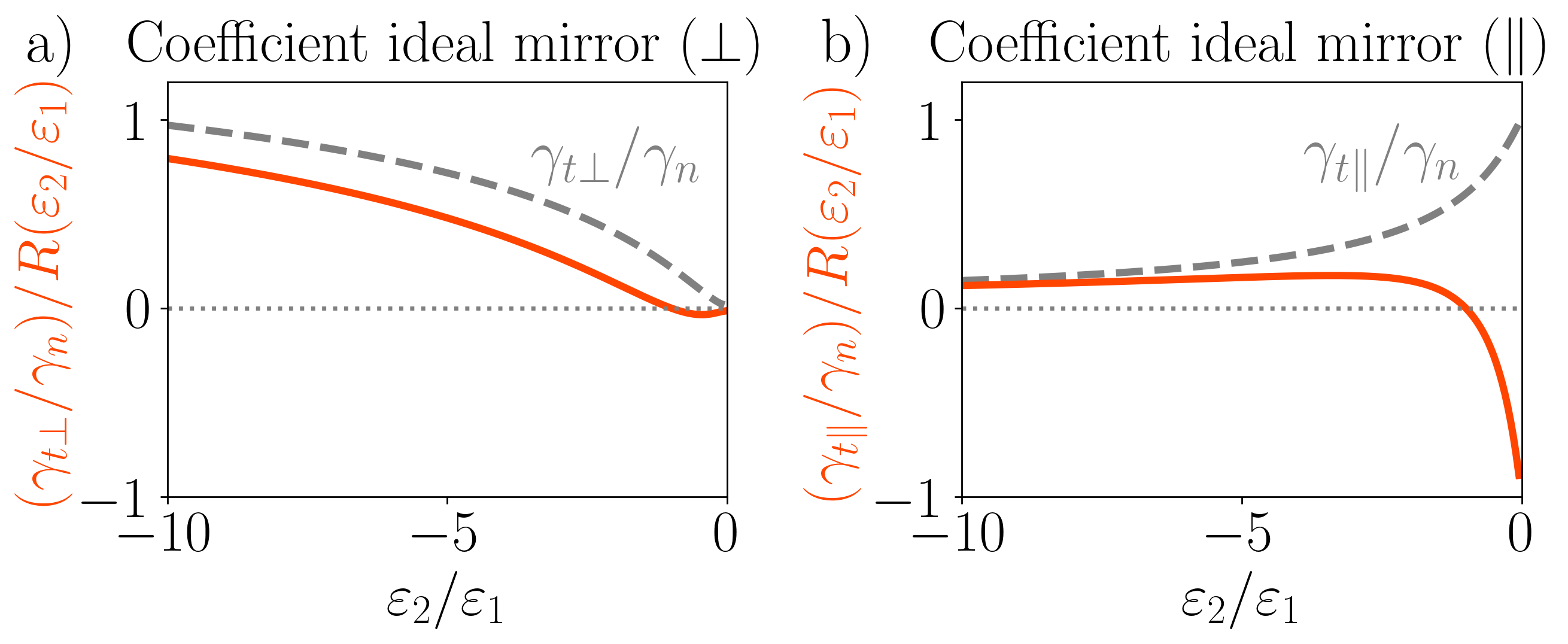}
\caption{(Color online) Solid (orange) lines correspond to the coefficients for determining the threshold amplitude given in Eq.~\eqref{eq_thresh_amp_IM} for the case of an plasmonic metal with $\text{Re}[\varepsilon_2] <0$ and $\text{Im}[\varepsilon_2] =0$. Panel a) corresponds to the case of a dipole perpendicular to the mirror and b) to the parallel case. The dashed (gray) lines correspond to the asymptotic values of the decay rates at short distances $k R_0 \rightarrow 0$ as a function of $\varepsilon_2 / \varepsilon_1$.}
\label{fig_coef_ideal_mirror}
\end{figure}

For an plasmonic metal where $\text{Im}[\varepsilon_2] = 0$ and $\text{Re}[\varepsilon_2] < 0$, the damping $\gamma_{t \perp, \parallel}$ becomes a constant independent of time that we have to calculate explicitly. The asymptotic values of the decay rates $\gamma_{t \perp, \parallel}$ are indicated by gray dashed lines in Figs.~\ref{fig_coef_ideal_mirror} a) and b). From the condition of amplification in Eq.~\eqref{eq_threshold_alpha} and the equation of the amplitude \eqref{eq_amplitude_modulation}, we find an equation similar to the case of a lossy mirror [see Eq.~\eqref{eq_thresh_amp_LM}]:
\begin{equation}\label{eq_thresh_amp_IM}
A_{\text{th} \perp, \parallel} = \frac{  \gamma_{t \perp, \parallel} / \gamma_n }{ \zeta_{\perp,\parallel} | R(\varepsilon_2/\varepsilon_1) | }.
\end{equation}
In this case, the Fresnel-like coefficient $R(\varepsilon_2 / \varepsilon_1)$ is positive for $\varepsilon_2 / \varepsilon_1 < -1$ but can become negative for $\varepsilon_2 / \varepsilon_1 > -1$ in which case the modulation acquires a phase shift. The threshold value in Eq.~\eqref{eq_thresh_amp_IM} is analogous to that of a perfect mirror [see Eqs.~\eqref{eq_a_perp_position} and \eqref{eq_a_parallel_position}]. The main difference lies in the denominator, which represents the dynamical Lamb shift, as no dynamical decay rate is present in this case. In fact, we can recover the perfect mirror case when $\varepsilon_2 \rightarrow - \infty$. In Figs.~\ref{fig_coef_ideal_mirror} a) and b), we show the behavior of the coefficients $(\gamma_{t \perp, \parallel }/\gamma_n)/R(\varepsilon_2 / \varepsilon_1)$ \mike{that determine the threshold amplitude} for the perpendicular and parallel case with solid orange lines. In the limit $\varepsilon_2 \rightarrow - \infty$, we recover the perfect mirror case. \textcolor{black}{For $\varepsilon_2 \rightarrow 0$, we obtain the asymptotic limits $(\gamma_{t \perp} / \gamma_n )\rightarrow 0$ and $(\gamma_{t \parallel} / \gamma_n )\rightarrow 1$. The surface behaves like a perfect mirror, but the phase of the reflected field is different from the perfect mirror case.}\\

\begin{figure}
\includegraphics[scale=0.3]{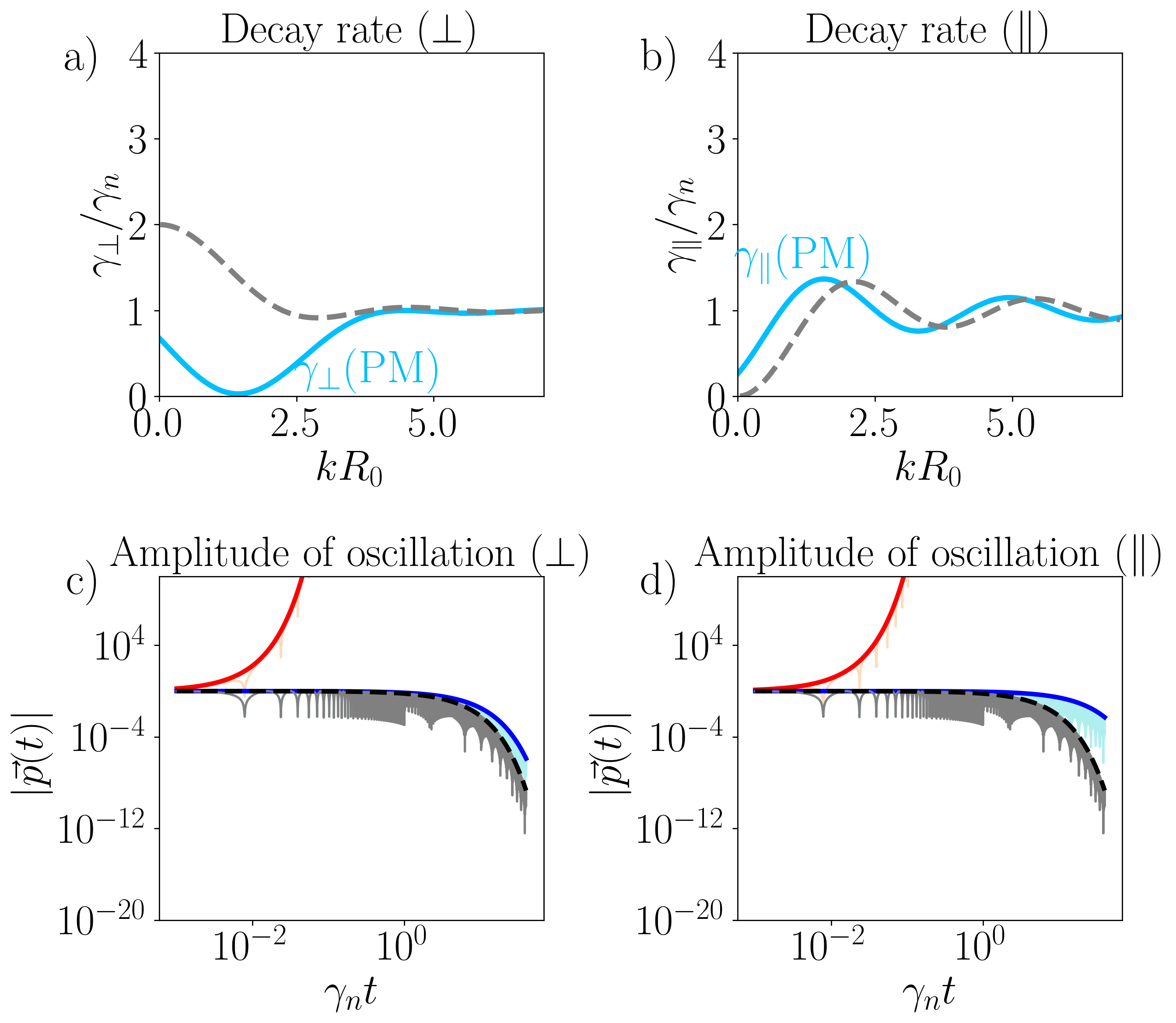}
\caption{(Color online) a), b) Decay rates for an plasmonic metal (PM) for the a) perpendicular and b) parallel orientations with $\varepsilon_2 / \varepsilon_1 = -4.5$. c), d) Qualitative behavior of the absolute value of the oscillations of a dipole on a logarithmic scale. We chose a distance given by $(k R_0)^{-1} = 12$ and the modulation amplitude is given by $A=0.67$. See the caption in Fig.~\ref{fig_decay_osc_LM}.}
\label{fig_decay_osc_IM}
\end{figure}

In Fig.~\ref{fig_decay_osc_IM}, we illustrate the distance-dependent behavior of the decay rates for a permittivity ratio of $\varepsilon_2 / \varepsilon_1 = -4.5$. In panels a) and b), we can see the constant limit of the decay rate at short distances $kR_0 \rightarrow 0$. Panels c) and d) display the behavior of the oscillation amplitude of the dipole, where the decay of $|\vec{p}(t)|$ in front of this particular plasmonic metal mirror is similar to the decay in homogeneous space, corresponding to an overlap between the two decreasing amplitudes.\\

\subsection{ENZ material}
\label{sec:enz}

In this section, we will analyze materials characterized by a permittivity $-1 \leq \text{Re}[\varepsilon_2 / \varepsilon_1] <0 $ and $\text{Im}[\varepsilon_2 / \varepsilon_1] > 0$, as indicated by the yellow dashed vertical line in Fig.~\ref{fig_real_imag_R} a). Let us first address the case where the Lamb shift vanishes $\text{Re}[R(\varepsilon_2 / \varepsilon_1)] = 0$, as shown by the dashed circular line in Fig.~\ref{fig_real_imag_R} a). Under these conditions, the threshold amplitude is given by
\begin{equation}\label{eq_amp_thresh_zero_Lamb}
A_{\text{th}\perp, \parallel} = \frac{2}{3}.
\end{equation}
In this case, the dynamic damping compensates the static decay allowing us to obtain amplification, which can also be seen from Eq.~\eqref{eq_thresh_amp_LM} given that $\text{Re}[R(\varepsilon_2 / \varepsilon_1)]=0$. While this threshold amplitude is not much smaller than one, the case remains of interest, as it captures the qualitative behavior of the system.\\

An interesting case \mike{within the ENZ regime} is the limit of $\varepsilon_2 / \varepsilon_1 \rightarrow -1$, \textcolor{black}{which corresponds to a plasmonic metal because of the vanishing imaginary part of the permittivity, as in the previous subsection}. In this case, the threshold has different values depending on the direction of the approach, as illustrated in Fig.~\ref{fig_thresh_ampl_LM} where different contour curves converge at this point, revealing a nonanalytic behavior of our approximation. In the short distance regime, setting $\text{Re}[\varepsilon_2 / \varepsilon_1] = -1$ and taking the limit $\text{Im}[\varepsilon_2 / \varepsilon_1]\rightarrow 0$, the threshold amplitude becomes
\begin{equation}
A_{\text{th}\perp, \parallel} = \frac{\sqrt{2}}{3}.
\end{equation}
On the other hand, when we approach $\varepsilon_2 / \varepsilon_1 \rightarrow -1$ using the curve where the Lamb shift contribution vanishes $\text{Re}[R(\varepsilon_2 / \varepsilon_1)]=0$, we obtain the same value of Eq.~\eqref{eq_amp_thresh_zero_Lamb}. The case of permittivities around this point can be treated with Eq.~\eqref{eq_thresh_amp_LM} when the effect of the imaginary part $\text{Im}[R(\varepsilon_2 / \varepsilon_1)]$ is significant, or can be treated as an plasmonic mirror, with Eq.~\eqref{eq_thresh_amp_IM} when $\text{Im}[R(\varepsilon_2 / \varepsilon_1)] \approx 0$.\\

\mike{A further noteworthy limiting} case corresponds to a perfectly absorbing medium where $\text{Re}[\varepsilon_2  ]=0$ and $\text{Im}[\varepsilon_2  ] \rightarrow \infty$. The Fresnel coefficient approaches $ R(\varepsilon_2 / \varepsilon_1) \approx 1 $, and the threshold amplitude of modulation reduces to
\begin{equation}
A_{\text{th} \perp, \parallel} = \frac{1}{\zeta_{\perp, \parallel}}.
\end{equation}

\section{Physical implementations}
\label{Sec_phys_implement}

In this section, we outline key considerations for the experimental implementation of the present results. These involve three main components: the emitter, the reflecting surface, and the mechanical modulation of their separation. A wide range of emitters has been studied in this context, including rare-earth ions such as $\mathrm{Eu}^{3+}$ \cite{Drexhage_influence_surface}, Rydberg atoms \cite{Hulet_spontaneous_Rydberg}, molecules, and NV centers \cite{Mohtashami_2013}. Advances in metamaterials and metasurfaces now allow unprecedented control over the permittivity of reflecting surfaces \cite{Li_metasurfaces, SUN2025145}. Nanometer-scale mechanical control is enabled by cavity optomechanics \cite{Aspelmeyer_cav_optomech}, though challenges remain in characterizing motion at the macroscopic scale (limited by thermal noise \cite{Murch_obs_backaction_ultrcold}) and fabricating structures at the microscopic scale. State-of-the-art mechanical modulation frequencies reach only a few GHz \cite{Aspelmeyer_cav_optomech}, placing stringent constraints on suitable emitters and surfaces that could be adequate for observing the phenomena of amplitude amplification in a dipole with time-dependent position.\\

One approach is the moving-mirror analogy proposed in Ref.~\cite{Belyanin_coherent_rad_moving_atom}, where molecules traveling atop a periodic grating effectively mimic mirror motion. Notably, transitions between neighboring high-lying Rydberg states occur in the GHz range \cite{Cohadon_moving_mirrors}, slightly overlapping with optomechanical capabilities. Microdisk resonators \cite{Ding_disk_resonator} could therefore serve as high-frequency mirrors for amplifying Rydberg-state emission. Another route, inspired by Ref.~\cite{Kiasat2023}, is to use optical forces to trap emitters near a surface. In the ENZ regime, equilibrium points with harmonic confinement can occur extremely close to the surface, as shown in Ref.~\cite{Kiasat2023}, where the trap frequency depends on the mass of the emitter. Optical trapping should be implemented with a laser polarized parallel to the surface, and emitters with two orthogonal dipole moments would allow selective analysis of the perpendicular component to the mirror. Another approach to modulating the dipole-mirror distance involves using surface transverse waves to create a moving grating on the surface of a reflecting surface \cite{Bigler_STW}, although the amplitudes of the waves might not be enough for practical implementation. Finally, let us comment that other \mike{related} types of time-dependent environments can be considered, where amplification can occur. For example, a photonic time-crystal with a perfect mirror, atoms coupled to waveguides simulating a 1D environment \mike{with long-range interactions} \cite{PhysRevA.90.012113}, and in general spin-boson models with time-dependent couplings \cite{PhysRevA.98.052129}.\\

\section{Final remarks}\label{Sec_fin_rem}

We have investigated the modification of the decay rate of an electric dipole in a time-varying electromagnetic environment. By deriving the equation of motion for the dipole oscillator in terms of the dyadic Green’s function, we established a general framework capable of incorporating the effects induced by motion of the dipole close to arbitrary reflecting surfaces. While the equation of motion is generally non-Markovian, requiring the knowledge of previous positions of the dipole, we find that for motion within a subwavelength distance to the mirror, the feedback from the electromagnetic bath becomes instantaneous. Then, a Markovian dynamics is recovered with the emergence of time-dependent damping and Lamb shift terms governed by the instantaneous position of the dipole, which determine the equation of a parametric oscillator. The dynamical coefficients depend on the properties of the dipole and the relative permittivity of the mirror. We identified specific sinusoidal trajectories that enable the dipole dynamics to be mapped onto a Mathieu equation. A general condition for achieving parametric amplification when the frequency of modulation is twice the frequency of the dipole was also derived for short distances to an imperfect mirror, where the free-space contribution to the damping can be neglected. In this regime, we found that amplification could be possible for any values of the permittivity of the mirror. Interestingly, in the \mike{ENZ} regime, we find that amplification can occur by means of an abrupt variation of the dynamical damping without the need of a dynamical Lamb shift. Our results highlight the versatility of light–matter interactions, showing how engineered time-dependent environments using mechanical motion can give rise to unusual dynamics, like amplification by means of a time-dependent damping.\\

This framework opens several promising directions for future exploration. One natural extension is to consider structured emitters, such as nanoplatelets, that host multiple dipole moments. This could enable intricate control over the coupling and dynamics among dipoles within a single emitter. \mike{While we have restricted ourselves to the weak excitation regime, another straightforward generalization is to incorporate two-level systems exhibiting saturation effects and nonlinear dynamics} \cite{Belyanin_coherent_rad_moving_atom}. Additionally, the use of image methods can be generalized to more complex geometries; for instance, dipole amplification within a corner reflector may be analyzed by superposing the effects of an infinite set of image dipoles generated by intersecting mirrors. Our approach could also be extended to account for retardation and relativistic effects, as well as regimes involving strong light–matter coupling where non-perturbative treatments become necessary. Importantly, practical implementations must consider the finite extent of real mirrors, where edge and corner scattering may introduce deviations from the idealized behavior predicted in this study when they become of the order of the wavelength of the dipole. Finally, another line of investigation could be to consider surfaces with different permeabilities, such as magnetic reflectors, which also exhibit a negative refractive index \cite{Rupin_Lifetime_negative_refrac_index}.\\

\section{Supplementary information}

List of variables, detailed derivation of the dipole’s equation of motion, approximations for the reflected electric field, \mike{the procedure for the mapping onto a Mathieu equation}, and plots of the threshold amplitudes of modulation at different distances.\\

\section{Acknowledgments}

This work was supported by the Air Force Office of Scientific Research
(AFOSR) through the Multi-University Research Initiative (MURI) program no.~FA9550-22-1-0317.

\bibliography{bibliography}
\bibliographystyle{apsrev4-1}

\newcommand{\apptoctype}{atoc} 

\onecolumngrid

\appendix

\counterwithout{equation}{section}

\setcounter{section}{0}
\setcounter{figure}{0}
\setcounter{table}{0}
\setcounter{equation}{0}

\renewcommand{\thesection}{S\arabic{section}}
\renewcommand{\thefigure}{S\arabic{figure}}
\renewcommand{\thetable}{S\arabic{table}}
\renewcommand{\theequation}{S\arabic{equation}}
\makeatletter
\let\old@section\section
\renewcommand{\section}{%
  \@ifstar{\old@section*}{\app@section}%
}
\newcommand{\app@section}[1]{%
  \old@section{#1}%
  \addcontentsline{\apptoctype}{section}{\protect\numberline{\thesection}#1}%
}
\makeatother

\renewcommand{\appendixname}{}
\newpage

\addtocontents{\apptoctype}{\protect\setcounter{tocdepth}{2}}

\begin{center}
{\large
Supplementary Information for\\[4pt]
\textbf{``Dynamical Drexhage Effect: Amplified Emission in Time-Modulated Electromagnetic Environments''}
}

\end{center}


\phantomsection
\providecommand{\contentsname}{Contents}
\renewcommand{\contentsname}{Appendix Contents}

\setcounter{tocdepth}{2}   

\makeatletter
\@starttoc{\apptoctype}
\makeatother



\section{List of variables}

\begin{table}[h!]
  \centering
  \begin{tabular}{||l|p{6cm}|p{6.5cm}||}
  \hline
    Symbol & Description &  Notes / Definition \\
    \hline
    $\omega_0$ & Resonance frequency of the dipole oscillator &  \\
    $m$ & Effective mass of dipole oscillator &  \\
    $q$ & Effective charge of the dipole oscillator &  \\
    $\vec{p}_{q}$ & Dipole moment vector &  $\vec{p}_{q} = \vec{n}_p q \sqrt{\hbar /(2 m \omega_0)}$\\
    $\vec{n}_p$ & Unit vector along dipole moment $\vec{p}$ & $\|\vec{n}_p\|=1$ \\
        $\overline{\overline{G}}(\vec{r},\vec{x},\omega)$ 
    & Dyadic Green's function 
    &  $\overline{\overline{G}}_0 (\vec{x},\vec{y}, \omega) + \overline{\overline{G}}_s (\vec{x},\vec{y}, \omega)$\\
    $\vec{h}(t)$ & Dipole position vector & For perpendicular motion: $(0,0,h(t))$ \\
    $h(t)$ & Instantaneous distance from dipole to mirror & $h(t) = R_0[1+\eta(t)]$ \\
    $R_0$ & Average distance to mirror &  \\
    $\eta(t)$ & Relative displacement modulation function & Dimensionless; $|\eta(t)| < 1$ \\
    $\gamma_n$ & Decay rate in a homogeneous medium with refractive index $n$ &  \\
    $\Delta \upsilon(\vec{h}(t),\vec{h}(t_{\text{prev}}))$ &    Non-Markovian dynamical Lamb shift &  \\
    $\Gamma(\vec{h}(t),\vec{h}(t_{\text{prev}}))$ & Non-Markovian dynamical decay rate &  \\
    $\Delta\nu(\vec{h}(t))$ & Markovian dynamical Lamb shift &  \\
    $\gamma(\vec{h}(t))$ & Markovian dynamical decay rate &  \\
    $\varepsilon_1$ & Permittivity of embedding medium & Considered real and positive here\\
    $\varepsilon_2$ & Complex permittivity of mirror & $\varepsilon_2 = \varepsilon_2' + i\varepsilon_2''$ \\
    $R(\varepsilon_2/\varepsilon_1)$ & Fresnel-like reflection coefficient & $R(\varepsilon_2/\varepsilon_1) = (\varepsilon_2 - \varepsilon_1)/(\varepsilon_2 + \varepsilon_1 )$ \\
    $K_{\perp,\parallel}$ & Lamb-shift coupling coefficients &  \\
    $b_{\perp,\parallel}$ & Damping coupling coefficients & \\
    $\gamma_{t\perp,\parallel}$ & Constant background damping terms &  \\
    $A$ & Modulation amplitude of $\eta(t)$ & \\
    $\alpha_{\perp,\parallel}$ & Parametric drive strength & Related to $A$ via Eq.~\eqref{eq_amplitude_modulation} \\
    $\Omega$ & Modulation frequency of $h(t)$ &  $\Omega=2\omega_0$ for parametric resonance \\
    $k$ & Wavenumber in the medium & $k = n\omega_0/c$ where $n=\sqrt{\varepsilon_1\mu_1}/\sqrt{\varepsilon_0\mu_0}$\\
    \hline
  \end{tabular}
  \caption{Most important symbols used throughout the main text and the SI, with short explanations and definitions.}
  \label{tab:defs}
\end{table}

\section{Hamiltonian}\label{subsec_hamiltonian}

In this section we introduce the Quantum Electrodynamics (QED) Hamiltonian that describes the dipole in non-relativistic motion in a general electromagnetic environment described by a dyadic Green's function. We use two formalisms: Standard QED and Macroscopic QED. The standard version is formulated in terms of electromagnetic modes for situations where there are no losses, while the macroscopic QED formalism is given in terms of a dyadic Green's function that can model lossy materials. We introduce the two formalisms to show explicitly that our resulting equation of motion for the dipole can be obtained from both formalisms.\\

The Hamiltonian of an electric dipole oscillator interacting with the electromagnetic field at a position $\vec{h}(t)$ has the form
\begin{equation}\label{eq_Hamiltonian_dipole_lin_coupling}
\hat{H} = \hbar \omega_0 ( \sigma^\dagger(t) \sigma(t) + 1/2) + \hat{H}_{r} - \hat{\vec{p}}(t)\cdot \hat{\vec{E}}(t), 
\end{equation}
where $\omega_0$ is the frequency of the dipole, $\hat{H}_r$ is the Hamiltonian of the free electromagnetic field, which for standard QED it is given by electromagnetic modes, or for macroscopic QED is given in terms of polaritonic modes, and we have also introduced a linear coupling between the dipole and the electric field. In standard QED the radiation Hamiltonian given by
\begin{equation}\label{eq_ham-free-normQED-SI}
\hat{H}_r = \sum_{\vec{k} \alpha} \hbar \omega_k a^{\dagger}_{\vec{k} \alpha}(t) a_{\vec{k} \alpha}(t),
\end{equation}
where we have defined the creation and annihilation operators of the electromagnetic field given by $a_{\vec{k}\alpha}^{\dagger}$ and $a_{\vec{k}\alpha}$, with $\vec{k}$ a wave vector, $\omega_k = c |\vec{k}|/n$ is the frequency of the wave vector and $\alpha$ is a polarization index \cite{book_Gerry}.  The local electric field operator that couples to the dipole at the instantaneous position $\vec{h}(t)$ is given by
\begin{equation}\label{eq_e-field-normal-SI}
\hat{\vec{E}}(t)= \sum_{\vec{k} \alpha} \Bigg( \frac{\hbar \omega_k}{2 \varepsilon_0} \Bigg)^{1/2} [ a_{\vec{k}\alpha}(t) \vec{F}_{\vec{k}\alpha}(\vec{h}(t)) + h.c],
\end{equation}
where the modes $ \vec{F}_{\vec{k}\alpha}(\vec{r}) $ are solutions to the vector wave equation \cite{book_Novotny},
\begin{equation}\label{vector_wave_equation}
\nabla \times \nabla \times \vec{F}_{\vec{k} \alpha}(\vec{r}) - k^2 \vec{F}_{\vec{k} \alpha}(\vec{r}) = \vec{0}.
\end{equation}
In general, the modes $\vec{F}_{\vec{k} \alpha}(\vec{r})$ are not easy to obtain for arbitrary geometries, except for some cases. As an example, for free space, the modes are the usual plane waves in a quantization volume $V$ and wave vector $\vec{k} = (k_x,k_y,k_z) $ \cite{book_Gerry}:
\begin{equation}\label{free_modes}
\vec{F}_{\vec{k} \alpha}(\vec{r}) = i \frac{\vec{e}_{\vec{k} \alpha} e^{i \vec{k} \cdot \vec{r} }}{\sqrt{V}}.
\end{equation}
where $\vec{e}_{\vec{k} \alpha}$ are polarization vectors. For a half-space filled with a perfect mirror, the electromagnetic modes are \cite{Arnoldus_George_spontaneous_decay}
\begin{equation}\label{eq_Drexhage_modes-Arnoldus-George}
\vec{F}_{\vec{k}\alpha}(\vec{r}) = \frac{\vec{e}_{\vec{k} \alpha} e^{i \vec{k} \cdot \vec{r}} +  \vec{\tilde{e}}_{\vec{k} \alpha} e^{i \vec{k}' \cdot \vec{r}} }{\sqrt{2V} },
\end{equation}
where the reflected wave vector is defined as $\vec{k}' = (k_x,k_y,-k_z)$. We have also introduced the $s$- and $p$-polarization vectors:
\begin{equation}\label{eq_polarization_vec_s}
\vec{e}_{\vec{k} s} = \frac{1}{k_{\parallel}} \vec{k}_{\parallel} \times \vec{e}_z,
\end{equation}
\begin{equation}\label{eq_polarization_vec_p}
\vec{e}_{\vec{k} p} = \frac{1}{k} \vec{k} \times \vec{e}_{\vec{k} s},
\end{equation}
where $\vec{e}_z$ is the unit vector perpendicular to the surface of the mirror and $k = |\vec{k}|, \;\; k_{\parallel} = |\vec{k}_{\parallel}|,  \;\; k_{\perp} = |\vec{k}_{\perp}|$ and $\vec{k}_{\parallel} = (k_x, k_y,0) $. We have also defined the reflected polarization vectors as
\begin{equation}\label{eq_reflect_polarization_vec}
\vec{\tilde{e}}_{\vec{k} \alpha} = \vec{e}_{\vec{k} \alpha}^{\perp} - \vec{e}_{\vec{k} \alpha}^{\parallel},
\end{equation}
where for a given polarization vector $\vec{e}_{\vec{k}\alpha} = ( e_{\vec{k}\alpha x} , e_{\vec{k}\alpha y}, e_{\vec{k}\alpha z} ) $ we have defined the perpendicular component to the mirror $\vec{e}_{\vec{k}\alpha}^{\perp} = ( 0, 0, e_{\vec{k}\alpha z} ) $ and the parallel component $\vec{e}_{\vec{k}\alpha}^{\parallel} = ( e_{\vec{k}\alpha x} , e_{\vec{k}\alpha y}, 0 ) $.\\

Differently, in macroscopic QED the radiation Hamiltonian is given by \cite{Chuang_TC_macroQED_perspective}
\begin{equation}\label{eq_Ham_macro_QED_SI}
\hat{H}_r = \int d^{3} \vec{r} \int_{0}^{\infty} d \omega \; \hbar \omega \hat{\vec{f}}^{\dagger}(\vec{r}, \omega) \cdot \hat{\vec{f}}(\vec{r}, \omega),
\end{equation}
where we have introduced the polaritonic modes that satisfy bosonic commutation relations $[f_{j}(\vec{y}, \nu), f_{l}^{\dagger}(\vec{x}, \omega)] = \delta_{j l} \delta (\vec{y} - \vec{x}) \delta (\nu - \omega)$ with $l$ and $j$ indices that indicate spatial components $x, y, z$. The electric field is given in terms of a general dyadic Green's function
\begin{equation}\label{eq_e_field_macroQED-SI}
\hat{\vec{E}} ( t) = i \sqrt{ \frac{\hbar}{\pi \varepsilon_0} } \int_{0}^{\infty} d \omega \int d^{3} \vec{x} \;\frac{\omega^2}{c^2} \sqrt{ \text{Im} [\varepsilon_r (\vec{x}, \omega) ] } \overline{\overline{G}} (\vec{h}(t) , \vec{x} , \omega) \hat{\vec{f}} (\vec{x} , \omega,t) + \text{h.c.},
\end{equation}
where $\varepsilon_r(\vec{x}, \omega)$ is the relative local permittivity, $\varepsilon_0$ the permittivity in free space, and $c$ is the speed of light in vacuum. Eqs. \eqref{eq_Ham_macro_QED_SI}-\eqref{eq_e_field_macroQED-SI} are more general than Eqs. \eqref{eq_ham-free-normQED-SI}-\eqref{eq_e-field-normal-SI} since they allow for the description of electromagnetic modes in complex absorptive media that might not be translationally invariant; under the limiting conditions of lossless media, where wavevectors are good mode labels, Eq. \eqref{eq_Ham_macro_QED_SI} yields the same physics as Eq. \eqref{eq_ham-free-normQED-SI}.

\section{Local electric field in terms of dyadic Green's function}\label{subsec_dyadic_green_e}

In this section, we derive the right hand side of Eq.~\eqref{eq_motion_dipole_dyadic} within standard QED, in terms of electromagnetic modes, as well as from macroscopic QED, in terms of polaritonic modes.

\subsection{Standard QED}

From the Hamiltonian given in Eq.~\eqref{eq_Hamiltonian_dipole_lin_coupling}, in terms of electromagnetic modes, we can obtain the equation of motion for the expectation value of the annihilation operator
\begin{equation}\label{eq_annih_field_Heisenberg}
i \hbar \frac{d}{dt} a_{\vec{l} \beta}(t) = \hbar \omega_l  a_{\vec{l}\beta}(t)  - \vec{p}(t)  \cdot \vec{F}_{\vec{l}\beta}^{*}(\vec{h}(t)) \Bigg(\frac{\hbar \omega_l}{2 \varepsilon_0}\Bigg)^{1/2},
\end{equation}
where, to simplify the notation, here and in the following, we omit the angle brackets of the expectation value and abbreviate $ \langle a_{\vec{l}\beta} (t) \rangle \rightarrow  a_{\vec{l}\beta} (t)$, and $\langle \hat{\vec{p}}(t) \rangle\rightarrow \vec{p}(t) $. The electric field is found by solving the Heisenberg differential equations. Hence, we can separate the field into an homogeneous solution $\vec{E}_{h}(t)$ and a particular solution $\vec{E}_{p}(t)$ such that $\vec{E}(t) = \vec{E}_{h}(t) + \vec{E}_{p}(t)$, where the first term corresponds to the field in the non-interacting case and the last one can be obtained by Fourier transforming Eq.~\eqref{eq_annih_field_Heisenberg},
\begin{equation}\label{eq_a_fourier_dipole}
 a_{\vec{l} \beta}  (\nu) = \frac{1 }{ \hbar \omega_l - \hbar \nu }  \int_{- \infty}^{\infty} d \omega_1\; \vec{p} (\omega_1) \;  \cdot \vec{\mathcal{F}}_{\vec{l} \beta}^{*}(  \omega_1 - \nu )  \Bigg(\frac{\hbar \omega_l}{2 \varepsilon_0}\Bigg)^{1/2},
\end{equation}
where $\vec{p} (\omega_1)$ is the Fourier transform of the dipole operator
\begin{equation}
\vec{p} (\omega) = \frac{1}{2 \pi} \int_{-\infty}^{\infty} d t\; e^{i \omega t} \vec{p}(t) ,
\end{equation}
and we have defined
\begin{equation}\label{eq_mode_fourier_t2}
\vec{\mathcal{F}}_{\vec{l}\beta}( \nu - \omega_1 ) =  \frac{1}{2 \pi} \int_{- \infty}^{\infty} d \tau\; e^{i (\nu - \omega_1) \tau } \vec{F}_{\vec{l}\beta}(\vec{h}(\tau)).
\end{equation}
The particular solution of the electric  field in frequency space is given by a Fourier transform
\begin{equation}\label{eq_e_field_nu}
\vec{E}_{p} (\nu) = \sum_{\vec{k} \sigma} \Bigg( \frac{\hbar \omega_k}{2 \varepsilon_0} \Bigg)^{1/2} \frac{1}{2 \pi} \int_{- \infty}^{\infty}d t\, e^{i \nu t} [  a_{\vec{k}\sigma}(t)  \vec{F}_{\vec{k}\sigma}(\vec{h}(t)) \, + \, \text{h.c.} ].
\end{equation}
We can use the convolution theorem to express the Fourier transform as an integral and we can notice that the electric field can be split into positive and negative frequency components $\vec{E}_{p}(\nu) = \vec{E}_{p}^{+}(\nu) + \vec{E}_{p}^{-}(\nu)$, where
\begin{equation}\label{eq_E_plus}
\vec{E}_{p}^{+}(\nu) = \sum_{\vec{k} \sigma} \Bigg( \frac{\hbar \omega_k}{2 \varepsilon_0} \Bigg)^{1/2}  \int_{- \infty}^{\infty} d \omega_1 \; a_{\vec{k}\sigma}(\omega_1) \vec{\mathcal{F}}_{\vec{k}\sigma}(\nu - \omega_1) 
\end{equation}
and
\begin{equation}\label{eq_E_minus}
\vec{E}_{p}^{-}(\nu) = \sum_{\vec{k} \sigma} \Bigg( \frac{\hbar \omega_k}{2 \varepsilon_0} \Bigg)^{1/2}    \int_{- \infty}^{\infty} d \omega_1 \; a^{\dagger}_{\vec{k}\sigma}(\omega_1) \vec{\mathcal{F}}_{\vec{k}\sigma}^{*}( \omega_1 - \nu)   . 
\end{equation}
It can be shown that the inverse Fourier transform of $\vec{E}_{p}^{-}(\nu)$ will have a zero contribution in time domain due to causality. After substituting Eqs.~\eqref{eq_a_fourier_dipole} and \eqref{eq_mode_fourier_t2} into Eq. \eqref{eq_E_plus}, we obtain
\begin{equation}
\vec{E}_{p}^{+}(\nu) = \sum_{\vec{k} \sigma} \Bigg( \frac{\hbar \omega_k}{2 \varepsilon_0} \Bigg)    \int_{- \infty}^{\infty}  d\omega_1   \frac{1 }{ \hbar \omega_k - \hbar \omega_1 }  \int_{- \infty}^{\infty} d \omega_2\;  \vec{p}(\omega_2)   \cdot  \vec{\mathcal{F}}_{\vec{k}\sigma}^{*}( \omega_2 - \omega_1 ) \vec{\mathcal{F}}_{\vec{k}\sigma}(\nu - \omega_1).
\end{equation}
We can inverse Fourier transform to obtain the expression in time domain and also introduce the modes using Eq.~\eqref{eq_mode_fourier_t2} to identify an expression that involves the dyadic Green's function. The electric field in time domain is
\begin{equation}
\begin{split}
\vec{E}_{p}(t) &= \int_{- \infty}^{\infty} d \nu\; e^{- i \nu t} \vec{E}_{p}^{+}(\nu)\\
&= \int_{- \infty}^{\infty} d \nu\; e^{- i \nu t} \sum_{\vec{k} \sigma} \Bigg( \frac{\hbar \omega_k}{2 \varepsilon_0} \Bigg)    \int_{- \infty}^{\infty}  d\omega_1\;    \frac{1 }{ \hbar \omega_k - \hbar \omega_1 }  \\
& \hspace{1cm} \times \int_{- \infty}^{\infty} d \omega_2\; \vec{p}(\omega_2) \;  \cdot   \frac{1}{2 \pi} \int_{- \infty}^{\infty} d \tau_1\; e^{i (\omega_1 - \omega_2) \tau_1 } \vec{F}_{\vec{k}\sigma}^{*}(\vec{h}(\tau_1)) \frac{1}{2 \pi} \int_{- \infty}^{\infty} d \tau_2\; e^{i (\nu - \omega_1) \tau_2 } \vec{F}_{\vec{k}\sigma}(\vec{h}(\tau_2)).   
\end{split}
\end{equation}
We can now use the Sokhotski–Plemelj theorem on the term $(\omega_k - \omega_1)^{-1}$ to obtain
\begin{equation}\label{eq_e_field_Dirac_principal_SI}
\begin{split}
\vec{E}_{p}(t) &= \int_{- \infty}^{\infty} d \nu\; e^{- i \nu t}        \int_{- \infty}^{\infty} d \omega_2\; \vec{p}(\omega_2) \;  \cdot   \frac{1}{2 \pi} \int_{- \infty}^{\infty} d \tau_1\;  \frac{1}{2 \pi} \int_{- \infty}^{\infty} d \tau_2   \\
& \hspace{1cm}  \times \int_{- \infty}^{\infty}  d\omega_1\;  e^{i (\omega_1 - \omega_2) \tau_1 }  e^{i (\nu - \omega_1) \tau_2 }  \sum_{\vec{k} \sigma} \Bigg( \frac{ \omega_k}{2 \varepsilon_0} \Bigg) \vec{F}_{\vec{k}\sigma}^{*}(\vec{h}(\tau_1)) \vec{F}_{\vec{k}\sigma}(\vec{h}(\tau_2))   \Bigg[ i \pi \delta (\omega_k - \omega_1) + \text{P}\Bigg(\frac{1}{\omega_k - \omega_1}\Bigg) \Bigg]  .
\end{split}
\end{equation}
Notice that the term $\vec{E}_{p}^{-}(\nu)$, see Eq.~\eqref{eq_E_minus}, would have a factor $\omega_k + \omega_1$ in the denominator, which would give a zero contribution when using the Sokhotski–Plemelj theorem because the frequencies $\omega_k$ are positive. A general solution for the annihilation operator would involve the homogeneous solution $a_0 e^{-i \omega_k t}$, which would also give us the homogeneous term of the electric field $\vec{E}_{h}(t)$. Since the expectation value of the annihilation operator vanishes in the vacuum,  $a_0 = 0$, we obtain $\vec{E}(t) = \vec{E}_{p}(t) $ and we will drop the subscript $p$ in the following. When we substitute the electric field in Eq.~\eqref{eq_e_field_Dirac_principal_SI} on the right hand side of Eq.\eqref{eq_motion_dipole_v1}, the driving term, we obtain
\begin{equation}\label{eq_driving_term_modes}
\begin{split}
\vec{p}_q \frac{2 \omega_0 \vec{p}_q}{\hbar} \cdot \vec{E}(t) &= \vec{p}_q \frac{2 \omega_0 }{\hbar} \int_{- \infty}^{\infty} d \nu \;e^{- i \nu t}        \int_{- \infty}^{\infty} d \omega_2  \;     \frac{1}{2 \pi} \int_{- \infty}^{\infty} d \tau_1\;  \frac{1}{2 \pi} \int_{- \infty}^{\infty} d \tau_2   \\
& \hspace{1cm}  \times \int_{- \infty}^{\infty}  d\omega_1\;  e^{i (\omega_1 - \omega_2) \tau_1 }  e^{i (\nu - \omega_1) \tau_2 } \\
& \; \times \sum_{\vec{k} \sigma} \Bigg( \frac{ \omega_k}{2 \varepsilon_0} \Bigg) \vec{p}(\omega_2) \cdot \vec{F}_{\vec{k}\sigma}^{*}(\vec{h}(\tau_1)) \vec{F}_{\vec{k}\sigma}(\vec{h}(\tau_2)) \cdot \vec{p}_q  \Bigg[ i \pi \delta (\omega_k - \omega_1) + \text{P}\Bigg(\frac{1}{\omega_k - \omega_1}\Bigg) \Bigg] . 
\end{split}
\end{equation}
We can factorize the expectation value of the dipole $\vec{p}(\omega_2) = \vec{p}_q (\sigma(\omega_2) + \sigma^{\dagger}(\omega_2))$ to move the frequency dependence of the dipole and introduce the outer product to get $ \vec{p}(\omega_2) \cdot \vec{F}_{\vec{k}\sigma}^{*}(\vec{h}(\tau_1)) \vec{F}_{\vec{k}\sigma}(\vec{h}(\tau_2)) \cdot \vec{p}_q = \vec{p}_q \cdot [ \vec{F}_{\vec{k}\sigma}^{*}(\vec{h}(\tau_1)) \otimes \vec{F}_{\vec{k}\sigma}(\vec{h}(\tau_2)) ] \cdot \vec{p} (\omega_2) $. In Eq.~\eqref{eq_driving_term_modes} we can find an expression similar to the dyadic Green's function \cite{book_Novotny},
\begin{equation}
\overline{\overline{G} } (\vec{r}, \vec{x} , \omega_1) = \sum_{\vec{k} \alpha} \frac{ \vec{F}^{*}_{\vec{k} \alpha}(\vec{x}) \otimes \vec{F}_{\vec{k} \alpha}(\vec{r})  }{2 \omega_1  \varepsilon \mu} \Bigg[ i \pi \delta (\omega_k - \omega_1) + \text{P} \Bigg(\frac{1}{\omega_k - \omega_1} \Bigg)  \Bigg].
\end{equation}
However, to completely identify the above identity, we can substitute the $\omega_k$ with $\omega_1$ in the numerator of Eq.~\eqref{eq_driving_term_modes} because the Dirac delta and principal value will evaluate the continuous functions at $\omega_1$ and we also have to multiply and divide by $ \mu_1 \omega_1$. Hence, we to obtain
\begin{equation}
\begin{split}
\vec{p}_q \frac{2 \omega_0 \vec{p}_q}{\hbar} \cdot \vec{E}(t) &= \vec{p}_q \frac{2 \omega_0 }{\hbar} \int_{- \infty}^{\infty} d \nu\; e^{- i \nu t}        \int_{- \infty}^{\infty} d \omega_2  \;     \frac{1}{2 \pi} \int_{- \infty}^{\infty} d \tau_1\;  \frac{1}{2 \pi} \int_{- \infty}^{\infty} d \tau_2   \\
& \hspace{1cm}  \times \int_{- \infty}^{\infty}  d \omega_1\;  e^{i (\omega_1 - \omega_2) \tau_1 }  e^{i (\nu - \omega_1) \tau_2 }   \mu_1 \omega_1^2   \vec{p}_q \cdot \overline{\overline{G} } (\vec{h}(\tau_2), \vec{h}(\tau_1) , \omega_1) \cdot \vec{p}(\omega_2).
\end{split}
\end{equation}
Performing the integral over $\nu$, which gives us a Dirac delta $\delta (t - \tau_2)$, it is easy to next evaluate the integral over $\tau_2$ to get
\begin{equation}
\begin{split}
\vec{p}_q \frac{2 \omega_0 \vec{p}_q}{\hbar} \cdot \vec{E}(t) &= \vec{p}_q \frac{2 \omega_0 }{\hbar}         \int_{- \infty}^{\infty} d \omega_2  \;    \frac{1}{2 \pi} \int_{- \infty}^{\infty} d \tau_1\;      \\
& \hspace{1cm}  \times \int_{- \infty}^{\infty}  d \omega_1\;  e^{i (\omega_1 - \omega_2) \tau_1 }  e^{ -i \omega_1 t }   \mu_1 \omega_1^2   \vec{p}_q \cdot \overline{\overline{G} } (\vec{h}( t ), \vec{h}(\tau_1) , \omega_1) \cdot \vec{p}(\omega_2).
\end{split}
\end{equation}
Finally, we identify the function $\vec{p}(\tau_1)$ as an inverse Fourier transform of the dipole given by the integral over $\omega_2$ to obtain
\begin{equation}
\begin{split}
\vec{p}_q \frac{2 \omega_0 \vec{p}_q}{\hbar} \cdot \vec{E}(t) &= \vec{p}_q \frac{2 \omega_0 }{\hbar}              \frac{1}{2 \pi} \int_{- \infty}^{\infty} d \tau_1\;      \\
& \hspace{1cm}  \times \int_{- \infty}^{\infty}  d \omega_1\;  e^{i \omega_1  (\tau_1 - t) }   \mu_1 \omega_1^2   \vec{p}_q \cdot \overline{\overline{G} } (\vec{h}( t ), \vec{h}(\tau_1) , \omega_1) \cdot \vec{p}(\tau_1).
\end{split}
\end{equation}
Rearranging terms, introducing the dipole moment $\vec{p}_q$ inside the integrals and factorizing $\vec{p}(\omega_2) = \vec{p}_q (\sigma(\omega_2) + \sigma^{\dagger}(\omega_2))$, we obtain Eq.~\eqref{eq_motion_dipole_dyadic} of the main text
\begin{equation}
\begin{split}
\vec{p}_q \frac{2 \omega_0 \vec{p}_q}{\hbar} \cdot \vec{E}(t) &=  \frac{2 \omega_0 }{\hbar}              \frac{1}{2 \pi} \int_{- \infty}^{\infty} d \tau_1\;      \int_{- \infty}^{\infty}  d \omega_1 \; e^{i \omega_1  (\tau_1 - t) }   \mu_1 \omega_1^2   \vec{p}_q \cdot \overline{\overline{G} } (\vec{h}( t ), \vec{h}(\tau_1) , \omega_1) \cdot \vec{p}_q \vec{p} (\tau_1).
\end{split}
\end{equation}

\subsection{Macroscopic QED}

In macroscopic QED, the Hamiltonian in Eq.~\eqref{eq_Hamiltonian_dipole_lin_coupling} is given in terms of polaritonic modes, which are vector operators. The equation of motion of the expectation value of the polaritonic mode is
\begin{equation}
i \hbar \frac{d}{dt} \vec{f}(\vec{y}, \nu , t) = \hbar \nu  \vec{f}(\vec{y}, \nu , t) + \Bigg(  i \sqrt{ \frac{\hbar}{\pi \varepsilon_0} } \Bigg)  \frac{\nu^2}{c^2} \sqrt{ \text{Im} [\varepsilon_r (\vec{y}, \nu) ] } \; \bigg[ \overline{\overline{G}}^{\, *} (\vec{h}(t) , \vec{y} , \nu) \bigg]^{T} \cdot \vec{p}(t) ,
\end{equation}
where we are omitting the brackets of the expectation value $\langle \hat{\vec{f}}(\vec{y}, \nu, t) \rangle \rightarrow \vec{f}(\vec{y}, \nu, t)$ and $\langle \hat{\vec{p}}(t) \rangle \rightarrow \vec{p}(t)$. We Fourier transform and solve for the polaritonic mode to obtain
\begin{equation}\label{eq_sol_polaritonic_mode_f}
\vec{f}(\vec{y}, \nu , \omega_t) =   \frac{1}{( \hbar \omega_t - \hbar \nu)} \Bigg(  i \sqrt{ \frac{\hbar}{\pi \varepsilon_0} } \Bigg)  \frac{\nu^2}{c^2} \sqrt{ \text{Im} [\varepsilon_r (\vec{y}, \nu) ] } \; \int_{- \infty}^{\infty} d \omega_1 \;\overline{\overline{ \mathcal{G} }}^{\dagger} ( \omega_1 - \omega_t, \vec{y} , \nu)  \cdot \vec{p}( \omega_1 ) ,
\end{equation}
where we have used the convolution theorem and have also defined the Fourier transform of the dyadic Green's function,
\begin{equation}\label{eq_G_normal_Fourier}
\overline{\overline{ \mathcal{G} }} (\omega_1, \vec{y} , \nu) = \frac{1}{2 \pi} \int_{- \infty}^{\infty} d \tau \; e^{i \omega_1 \tau} \;  \overline{\overline{G}} (\vec{h}( \tau ) , \vec{y} , \nu) .
\end{equation}
In the following, we will calculate the Fourier transform of the electric field evaluated in a time-dependent position given by macroscopic QED and substitute the polaritonic mode obtained by a Fourier transform. This field corresponds to a particular solution of the equations of motion so we will denote it by $\vec{E}_{p}(\nu)$. We can split it into positive and negative frequencies as
\begin{equation}
\vec{E}_{p}(\nu) = \vec{E}_{p}^{+}(\nu) + \vec{E}_{p}^{-}(\nu) ,
\end{equation}
where
\begin{equation}\label{eq_ep_plus_mcqed}
\begin{split}
\vec{E}_{p}^{+}(\nu) &= \frac{1}{2 \pi} \int_{- \infty}^{\infty} d t \; e^{i \nu t}   i \sqrt{ \frac{\hbar}{\pi \varepsilon_0} } \int_{0}^{\infty} d \omega \int d^{3} \vec{x} \;\frac{\omega^2}{c^2} \sqrt{ \text{Im} [\varepsilon_r (\vec{x}, \omega) ] } \overline{\overline{G}} (\vec{h}(t) , \vec{x} , \omega) \vec{f} (\vec{x} , \omega,t) 
\end{split}
\end{equation}
and
\begin{equation}
\begin{split}
\vec{E}_{p}^{-}(\nu) &= \frac{1}{2 \pi} \int_{- \infty}^{\infty} d t\;  e^{i \nu t} \Bigg[ - i \sqrt{ \frac{\hbar}{\pi \varepsilon_0} } \int_{0}^{\infty} d \omega \int d^{3} \vec{x} \frac{\omega^2}{c^2} \sqrt{ \text{Im} [\varepsilon_r (\vec{x}, \omega) ] } \vec{f}^{\dagger} (\vec{x} , \omega,t) \overline{\overline{G}}^{\dagger} (\vec{h}(t) , \vec{x} , \omega)     \Bigg].
\end{split}
\end{equation}
We now can focus on the term $ \vec{E}_{p}^{+}(\nu)$ that involves the annihilation operator $\vec{f}(\vec{x}, \omega, t)$ because the other term will become zero due to causality. Using the convolution theorem, we can express the Fourier transform of the mode and the dyadic Green's function evaluated at a time-dependent position
\begin{equation}\label{eq_convolution_GF_mcqed}
\frac{1}{2 \pi} \int_{- \infty}^{\infty} d t\;  e^{i \nu t} \overline{\overline{G}} (\vec{h}(t) , \vec{x} , \omega) \hat{\vec{f}} (\vec{x} , \omega,t) = \int_{- \infty}^{\infty} d \omega_1\; \overline{\overline{ \mathcal{G} }} (\nu - \omega_1, \vec{x} , \omega) \hat{\vec{f}}(\vec{x}, \omega, \omega_1).
\end{equation}
Substituting Eqs.~\eqref{eq_sol_polaritonic_mode_f} and \eqref{eq_convolution_GF_mcqed} into Eq.~\eqref{eq_ep_plus_mcqed}, we obtain
\begin{equation}
\begin{split}
\vec{E}_{p}^{+}(\nu) &=    \frac{1}{\pi \varepsilon_0}  \int_{0}^{\infty} d \omega  \Bigg( \frac{\omega^2}{c^2} \Bigg)   \int_{- \infty}^{\infty} d \omega_1  \;\frac{1}{2 \pi} \int_{- \infty}^{\infty} d \tau_1 \; e^{i (\nu - \omega_1) \tau_1}  \\
& \hspace{1cm} \times \frac{1}{(  \omega -  \omega_1)}     \; \int_{- \infty}^{\infty} d \omega_2\; \frac{1}{2 \pi} \int_{- \infty}^{\infty} d \tau_2 \; e^{i (\omega_1 - \omega_2) \tau_2 } \\
& \hspace{1cm} \times \int d^{3} \vec{x}\;  \frac{\omega^2}{c^2} \text{Im} [\varepsilon_r (\vec{x}, \omega) ] \overline{\overline{G}} (\vec{h}( \tau_1 ) , \vec{x} , \omega)  \overline{\overline{G}}^{\, \dagger} (\vec{h}( \tau_2 ) , \vec{x} , \omega)   \cdot \vec{p}( \omega_2 ).
\end{split}
\end{equation}
We next perform the integral over $\vec{x}$ using the following identity \cite{Knoll_QED_disp_abs_med}:
\begin{equation}\label{eq_identity_dyadic_epsilon_mqed}
\int d^{3} \vec{x}\;  \frac{\omega^2}{c^2} \text{Im}[\varepsilon (\vec{x},\omega)] \overline{\overline{G}} (\vec{r},\vec{x}, \omega) \overline{\overline{G}}^{\dagger} (\vec{y}, \vec{x}, \omega) = \text{Im} [\overline{\overline{G}} (\vec{r}, \vec{y} , \omega) ],
\end{equation}
to obtain
\begin{equation}\label{eq_Ep_plus_omega}
\begin{split}
\vec{E}_{p}^{+}(\nu) &=    \frac{1}{\pi \varepsilon_0}  \int_{0}^{\infty} d \omega\;   \Bigg( \frac{\omega^2}{c^2} \Bigg)   \int_{- \infty}^{\infty} d \omega_1  \frac{1}{2 \pi} \int_{- \infty}^{\infty} d \tau_1 \; e^{i (\nu - \omega_1) \tau_1}  \\
& \hspace{1cm} \times \frac{1}{(  \omega -  \omega_1)}     \; \int_{- \infty}^{\infty} d \omega_2\; \frac{1}{2 \pi} \int_{- \infty}^{\infty} d \tau_2 \; e^{i (\omega_1 - \omega_2) \tau_2 } \; \text{Im} \big[ \overline{\overline{G}}(\vec{h}( \tau_1 ), \vec{h}( \tau_2 ), \omega)  \big]  \cdot \vec{p}( \omega_2 ).
\end{split}
\end{equation}
The Sokhotski–Plemelj theorem allows us to introduce a Dirac delta and a principal value
\begin{equation}
\frac{1}{  \omega -  \omega_1 } = i \pi \delta (\omega - \omega_1) + \text{P}\Bigg(\frac{1}{\omega - \omega_1}\Bigg).
\end{equation}
The integral of the term that involves the Dirac delta $\delta (\omega - \omega_1)$ is straightforward to evaluate. The integral that involves the principal value $\text{P}(1/(\omega - \omega_1))$ can be evaluated using the Kramers–Kronig relations:
\begin{equation}
\int_{0}^{\infty} d \omega  \Bigg( \frac{\omega^2}{c^2} \Bigg) \text{P}\Bigg(\frac{1}{\omega - \omega_1}\Bigg) \text{Im} \big[ \overline{\overline{G}}(\vec{h}( \tau_1 ), \vec{h}( \tau_2 ), \omega)  \big] = \pi  \Bigg( \frac{\omega_1^2}{c^2} \Bigg) \text{Re} \big[ \overline{\overline{G}}(\vec{h}( \tau_1 ), \vec{h}( \tau_2 ), \omega_1)  \big].
\end{equation}
After performing the integral over $\omega$ in Eq.~\eqref{eq_Ep_plus_omega}, we obtain
\begin{equation}
\begin{split}
\vec{E}_{p}^{+}(\nu) &=    \frac{1}{  \varepsilon_0}     \int_{- \infty}^{\infty} d \omega_1\;  \frac{1}{2 \pi} \int_{- \infty}^{\infty} d \tau_1 \; e^{i (\nu - \omega_1) \tau_1}   \\
& \hspace{1cm} \times   \; \int_{- \infty}^{\infty} d \omega_2 \;\frac{1}{2 \pi} \int_{- \infty}^{\infty} d \tau_2 \; e^{i (\omega_1 - \omega_2) \tau_2 } \;  \Bigg( \frac{\omega_1^2}{c^2} \Bigg) \Bigg[ i \text{Im} \big[ \overline{\overline{G}}(\vec{h}( \tau_1 ), \vec{h}( \tau_2 ), \omega_1)  \big] + \text{Re} \big[ \overline{\overline{G}}(\vec{h}( \tau_1 ), \vec{h}( \tau_2 ), \omega_1)  \big] \Bigg]   \cdot \vec{p}( \omega_2 ).
\end{split}
\end{equation}
Inside the brackets, we can identify the dyadic Green's function $\overline{\overline{G}}(\vec{h}( \tau_1 ), \vec{h}( \tau_2 ), \omega_1)$ separated into real and imaginary part to get the electric field in time domain,
\begin{equation}
\begin{split}
\vec{E}(t) &= \int_{- \infty}^{\infty} d \nu\; e^{- i \nu t} \vec{E}_{p}^{+}(\nu)\\
&= \int_{- \infty}^{\infty} d \nu\; e^{- i \nu t}  \frac{1}{  \varepsilon_0}     \int_{- \infty}^{\infty} d \omega_1  \frac{1}{2 \pi} \int_{- \infty}^{\infty} d \tau_1 \; e^{i (\nu - \omega_1) \tau_1}   \\
& \hspace{1cm} \times      \; \int_{- \infty}^{\infty} d \omega_2 \;\frac{1}{2 \pi} \int_{- \infty}^{\infty} d \tau_2 \; e^{i (\omega_1 - \omega_2) \tau_2 } \;   \Bigg( \frac{\omega_1^2}{c^2} \Bigg) \overline{\overline{G}}(\vec{h}( \tau_1 ), \vec{h}( \tau_2 ), \omega_1)  \cdot \vec{p}( \omega_2 ).
\end{split}
\end{equation}
Integrating over $\nu$, we obtain a Dirac delta $\delta (t - \tau_1)$. Then, it is straightforward to evaluate the integral over $\tau_1$ to obtain
\begin{equation}\label{eq_almost_final_e-field-normalQED_SI}
\begin{split}
\vec{E}(t)  &= \frac{1}{  \varepsilon_0}     \int_{- \infty}^{\infty} d \omega_1    \; e^{-i   \omega_1 t}        \; \int_{- \infty}^{\infty} d \omega_2 \frac{1}{2 \pi} \int_{- \infty}^{\infty} d \tau_2 \; e^{i (\omega_1 - \omega_2) \tau_2 } \;   \Bigg( \frac{\omega_1^2}{c^2} \Bigg) \overline{\overline{G}}(\vec{h}( t ), \vec{h}( \tau_2 ), \omega_1)  \cdot \vec{p}( \omega_2 ).
\end{split}
\end{equation}
The integral over $\omega_2$ is an inverse Fourier transform that allows us to obtain $\vec{p}(\tau_2)$. Then, we can substitute the expression for the electric field into Eq.~\eqref{eq_motion_dipole_v1} to obtain Eq.~\eqref{eq_motion_dipole_dyadic}. Being more general than standard QED, the dyadic Green's function can model environments with complex permittivities, that can represent lossy materials, while in standard QED the modes are obtained for plasmonic metals in idealized translationally invariant environments where wavevectors are good mode labels \cite{Arnoldus_quantum_theo_fluo}.\\

\section{Equation of motion with a static position}\label{subsec_eq_motion_static_field}

In this section, we demonstrate that Eq.~\eqref{eq_motion_dipole_dyadic} reduces to the well-known dynamics of a dipole with a fixed position. We can consider a general position vector $\vec{h} = (h_x,h_y,h_z)$ such that we can integrate over $\tau_1$ on the right-hand side of Eq.~\eqref{eq_motion_dipole_dyadic} to get a Dirac delta,
\begin{equation}\label{eq_e-field_fixed_pos_SI}
\begin{split}
\vec{p}_{q} \frac{2 \omega_0 \vec{p}_{q}}{\hbar} \cdot \hat{\vec{E}}(t) =& \frac{1}{2\pi} \int_{- \infty}^{\infty}  d\omega_1   \int_{-\infty}^{\infty} d \tau_1   \;      e^{-i \omega_1  ( t - \tau_1) }      \Bigg( \frac{2 \omega_0 }{ \hbar } \Bigg) [ \mu_1 \omega_1^2]  \Bigg[   [\vec{p}_{q} \cdot  \overline{\overline{G}}(\vec{h}, \vec{h} , \omega_1) \cdot \vec{p}_{q}]  \vec{p}(\tau_1)   \Bigg]\\\\
=&  \int_{- \infty}^{\infty}  d\omega_1      \;      e^{-i \omega_1   t  }      \Bigg( \frac{2 \omega_0 }{ \hbar } \Bigg) [ \mu_1 \omega_1^2]  \Bigg[   [\vec{p}_{q} \cdot  \overline{\overline{G}}(\vec{h}, \vec{h} , \omega_1) \cdot  \vec{p}_{q}] \vec{p}(\omega_1)   \Bigg],
\end{split}
\end{equation}
where we identified the Fourier transform of the dipole
\begin{equation}
\vec{p}(\omega_1) = \frac{1}{2\pi}\int_{-\infty}^{\infty} d \tau_1\;  e^{i \omega_1 \tau_1} \vec{p}(\tau_1).
\end{equation}
We see that the right-hand side of Eq.~\eqref{eq_e-field_fixed_pos_SI} is proportional to the Fourier transform of the electric field of the dipole at position $\vec{h}$ in terms of the dyadic Green's function $\vec{E}(\omega_1) = \mu_1 \omega_1^2 \overline{\overline{G}}( \vec{h}, \vec{h}, \omega_1 ) \cdot \vec{p}(\omega_1)$ \cite{book_Novotny}. Then, we obtain the well-known decay rate and Lamb shift by using the Born-Markov approximation, where we approximate $\vec{p}(\omega_1) \approx \vec{p}(0) \delta (\omega_0 - \omega_1)$, as in Ref.~\cite{Chance_lifetime_emitting_molecule}. We obtain
\begin{equation}
\begin{split}
\vec{p}_{q} \frac{2 \omega_0 \vec{p}_{q}}{\hbar} \cdot \hat{\vec{E}}(t) =&      \Bigg( \frac{2 \omega_0 }{ \hbar } \Bigg) [ \mu_1 \omega_0^2]     [\vec{p}_{q} \cdot  \overline{\overline{G}}(\vec{h}, \vec{h} , \omega_0) \cdot  \vec{p}_{q}] \vec{p}(0) e^{-i \omega_0   t  } \\\\
=& \omega_0 [ i \gamma(\vec{h}) + \Delta \nu (\vec{h}) ] \vec{p}(0) e^{- i \omega_0 t}
\end{split}
\end{equation}
where we have separated the dyadic Green's function into real and imaginary part and introduced the decay and Lamb shift coefficients, given in Eqs.~\eqref{eq_aux_decay_rates_imprf} and \eqref{eq_aux_Lamb_shift_imprf}. In the Born-Markov approximation, we can approximate $i \omega_0 \vec{p}(0) e^{-i \omega_0 t} \approx - d\vec{p} / dt $. Hence, in the weak-interaction regime, the retardation effects give rise to a friction and frequency shift,
\begin{equation}\label{eq_dec_lamb_static_SI}
\begin{split}
\vec{p}_{q} \frac{2 \omega_0 \vec{p}_{q}}{\hbar} \cdot \hat{\vec{E}}(t) =&  - \gamma(\vec{h}) \frac{d \vec{p}(t)}{dt} + \omega_0 \Delta \nu (\vec{h}) \vec{p}(t).
\end{split}
\end{equation}
Another route to obtain this Markovian equation is to use the convolution theorem on the right-hand side of Eq.~\eqref{eq_e-field_fixed_pos_SI} to obtain a integral over time,
\begin{equation}\label{eq_kernel_p_SI}
\vec{p}_{q} \frac{2 \omega_0 \vec{p}_{q}}{\hbar} \cdot \hat{\vec{E}}(t) =  \frac{1}{2 \pi} \int_{-\infty}^{\infty} d \tau\; \overline{\overline{K}}(\vec{h},\vec{h}, t - \tau) \vec{p}(\tau),
\end{equation}
where
\begin{equation}
\overline{\overline{K}}(\vec{h},\vec{h}, t - \tau) = \int_{- \infty}^{\infty}  d\omega_1      \;      e^{-i \omega_1   (t - \tau)  }      \Bigg( \frac{2 \omega_0 }{ \hbar } \Bigg) [ \mu_1 \omega_1^2]    [\vec{p}_{q} \cdot  \overline{\overline{G}}(\vec{h}, \vec{h} , \omega_1) \cdot  \vec{p}_{q}]
\end{equation}
and
\begin{equation}\label{eq_p_tau_evolution_Born}
\vec{p}(\tau) = \int_{- \infty}^{\infty}  d\omega_1      \;      e^{-i \omega_1   \tau  } \vec{p}(\omega_1).
\end{equation}
If we use the time-domain Born approximation $\vec{p}(t) \approx \vec{p}(0) e^{- i \omega_0 t}$ in Eq.~\eqref{eq_kernel_p_SI}, we can obtain a factor that depends on the present time $t$,
\begin{equation}
\vec{p}_{q} \frac{2 \omega_0 \vec{p}_{q}}{\hbar} \cdot \hat{\vec{E}}(t) =  \frac{1}{2 \pi} \int_{-\infty}^{\infty} d \tau\; \overline{\overline{K}}(\vec{h},\vec{h}, t - \tau) \vec{p}(0) e^{- i \omega_0 t}.
\end{equation}
Using a change of variables $s = t- \tau$ and identifying the Fourier transform of $\overline{\overline{K}}(\vec{h},\vec{h}, s)$, we arrive at the same result given in Eq.~\eqref{eq_e-field_fixed_pos_SI}. The key point is that, in the weak-interaction regime, the evolution of the dipole and the effect of the electromagnetic bath are dominated by the frequency component $\omega_0$. Then, in Eq.~\eqref{eq_e-field_fixed_pos_SI} we can also perform the integral over $\omega_1$ and use the Born approximation in time domain to obtain Eq.~\eqref{eq_e-field_fixed_pos_SI} and, hence, obtain the decay rate and the Lamb shift, as in Eq.~\eqref{eq_dec_lamb_static_SI}. The main difficulty is that the integral over $\omega_1$ in Eq.~\eqref{eq_e-field_fixed_pos_SI} is not easy to perform, particularly, for the case of lossy mirrors. However, we can obtain asymptotic expressions for the decay rate and Lamb shift for distances close to the reflecting surface as in Ref. \cite{Kocharovsky_reversal_of_radiation_reaction}.

\section{Reflected electric field of a moving dipole}\label{subsec_ref_elec_field_perf}

This section provides a detailed calculation of the electric field generated by a moving dipole near a non-dispersive mirror that fills half-space $z<0$ with constant permittivity $\varepsilon_{2}$ and a constant permeability $\mu_{2}$. The medium that surrounds the dipole has corresponding constants $\varepsilon_{1}$ and $\mu_{1}$. For simplicity, we consider trajectories perpendicular to the mirror given by $\vec{h}(t) = (0,0, h(t))$.\\

The dyadic Green's function that represents an infinite reflecting surface can be written as the sum of a free space dyadic Green's function $\overline{\overline{G}}_{0}(\vec{x},\vec{y},\omega)$ and the one that generates the scattered electric field $\overline{\overline{G}}_{s}(\vec{x},\vec{y},\omega)$ \cite{book_Novotny}, such that the total driving term of the dipole in Eqs.~\eqref{eq_motion_dipole_v1} or \eqref{eq_motion_dipole_dyadic} can be written as
\begin{equation}\label{eq_driving_fields}
 \vec{p}_{q} \frac{2 \omega_0 \vec{p}_{q}}{\hbar} \cdot \vec{E}(t) = \vec{p}_{q} \frac{2 \omega_0 \vec{p}_{q}}{\hbar} \cdot  \vec{E}_{0}(t) + \vec{p}_{q} \frac{2 \omega_0 \vec{p}_{q}}{\hbar} \cdot  \vec{E}_{s}(t),
\end{equation}
where the first term corresponds to the free-space contribution, and the last one to the reflected electric field. Even under modulation of the position of the dipole, the free space contribution gives a constant decay. Hence, in the rest of this section we will focus on the scattered electric field $\vec{p}_{q} 2 \omega_0 \hbar^{-1} \vec{p}_{q} \cdot  \vec{E}_{s}(t)$ of Eq.~\eqref{eq_driving_fields}, and at the end we will point out how to obtain the constant damping that comes from the free-space dyadic Green's function.\\

The driving term that involves the scattered electric field is given in terms of the corresponding dyadic Green's function by
\begin{equation}\label{eq_scattered_e_field_supplementary}
\frac{\vec{p}_{q} 2 \omega_0 \vec{p}_{q} }{\hbar}  \cdot  \vec{E}_{s}(t) = \frac{1}{2\pi} \int_{- \infty}^{\infty}  d\omega_1   \int_{-\infty}^{\infty} d \tau_1   \;      e^{-i \omega_1  ( t - \tau_1) }     \Bigg( \frac{2 \omega_0 }{ \hbar } \Bigg) [ \mu_1 \omega_1^2]  \Bigg[   [\vec{p}_{q} \cdot  \overline{\overline{G}}_{s}(\vec{h}(t), \vec{h}( \tau_1  ) , \omega_1) \cdot \vec{p}_{q}]  \vec{p}(\tau_1)  \Bigg].
\end{equation}
For convenience, let us define the dot product between the dipole moment and an auxiliary electric field:
\begin{equation}\label{eq_elec_field_delta_dipole}
\vec{p}_{q} \cdot \vec{E}_{\text{aux}} ( \vec{h}(t), \vec{h}( \tau_1  ) , t - \tau_1 ) = \vec{p}_{q} \cdot \int_{- \infty}^{\infty}  d\omega_1 \;      e^{-i \omega_1 (t - \tau_1) }   \mu_1 \omega_1^2         [  \overline{\overline{G}}_{s}(\vec{h}(t), \vec{h}( \tau_1  ) , \omega_1) \cdot \vec{p}_q],
\end{equation}
such that Eq.~\eqref{eq_scattered_e_field_supplementary} can be written as
\begin{equation}\label{eq_es_field_supplementary}
\frac{\vec{p}_{q} 2 \omega_0 \vec{p}_{q} }{\hbar}  \cdot  \vec{E}_{s}(t) = \frac{1}{2\pi}  \int_{-\infty}^{\infty} d \tau_1   \;   \Bigg( \frac{2 \omega_0 }{ \hbar } \Bigg)   [   \vec{p}_{q} \cdot \vec{E}_{\text{aux}} ( \vec{h}(t), \vec{h}( \tau_1  ) , t - \tau_1 )   ]  \vec{p}(\tau_1).
\end{equation}
The electric field $\vec{E}_{\text{aux}}$ in Eq.~\eqref{eq_elec_field_delta_dipole} corresponds to the one generated by a frequency-independent dipole $\vec{p}_q$. Hence, in time-domain the field $\vec{E}_{\text{aux}}$ corresponds to the one generated by a Dirac delta pulse. For a general lossy mirror we cannot obtain an explicit expression for $\vec{E}_{\text{aux}} ( \vec{h}(t), \vec{h}( \tau_1  ) , t - \tau_1 )$. However, we can argue that it will depend on a Dirac delta, and its derivatives, evaluated at a time in the past $\delta (t - \tau_1 -n |h(t) + h(\tau_1)| /c)$, with $n = \sqrt{\varepsilon_1 \mu_1} / \sqrt{\varepsilon_0 \mu_0} $ the refractive index. This can be seen from the general structure of the dyadic Green's function of the scattered field \cite{book_Novotny}:
\begin{equation}\label{eq_scattered_dyadic_SI}
\overline{\overline{G}}_{s}(\vec{h}, \vec{r}, \omega_1) = \frac{i}{8 \pi k_1^2} \int_{0}^{\infty}  dk_{\rho} \frac{k_{\rho}}{ \sqrt{k_{1}^2 - k_\rho^2}  }  \begin{pmatrix}
k_1^2 r^s - (k_{1}^2 - k_\rho^2) r^p & 0 & 0\\
0 & k_1^2 r^s - (k_{1}^2 - k_\rho^2) r^p & 0 \\
0 & 0 & 2 k_{\rho}^2 r^{p}
\end{pmatrix}
e^{i \sqrt{k_{1}^2 - k_\rho^2}  (h_z + r_z)} ,
\end{equation}
where $k_1 = \omega_1 n /c$ is the wave vector and we have introduced the Fresnel coefficients of $p$ polarization
\begin{equation}
r^p = \frac{\varepsilon_2 \sqrt{ \frac{\omega^2_0}{c^2} \frac{\varepsilon_1 \mu_1 }{\varepsilon_0 \mu_0} - k_{\rho}^2 }  - \varepsilon_1 \sqrt{ \frac{\omega^2_0}{c^2} \frac{\varepsilon_2 \mu_2 }{\varepsilon_0 \mu_0}  - k_{\rho}^2} }{ \varepsilon_2 \sqrt{ \frac{\omega^2_0}{c^2} \frac{\varepsilon_1 \mu_1 }{\varepsilon_0 \mu_0} - k_{\rho}^2 }  + \varepsilon_1 \sqrt{ \frac{\omega^2_0}{c^2} \frac{\varepsilon_2 \mu_2 }{\varepsilon_0 \mu_0}  - k_{\rho}^2}  },
\end{equation}
and the corresponding $s$ polarization coefficient
\begin{equation}
r^s = \frac{ \sqrt{ \frac{\omega^2_0}{c^2} \frac{\varepsilon_1 \mu_1 }{\varepsilon_0 \mu_0} - k_{\rho}^2 }  -  \sqrt{ \frac{\omega^2_0}{c^2} \frac{\varepsilon_2 \mu_2 }{\varepsilon_0 \mu_0}  - k_{\rho}^2} }{  \sqrt{ \frac{\omega^2_0}{c^2} \frac{\varepsilon_1 \mu_1 }{\varepsilon_0 \mu_0} - k_{\rho}^2 }  +  \sqrt{ \frac{\omega^2_0}{c^2} \frac{\varepsilon_2 \mu_2 }{\varepsilon_0 \mu_0}  - k_{\rho}^2}  }.
\end{equation}
Let us focus on the exponential factor $e^{i \sqrt{k_{1}^2 - k_\rho^2}  (h_z + r_z)}$ of the dyadic Green's function. Upon integration over the propagating modes $k_{\rho} < k_1$ it gives us an exponential factor $e^{i \omega_1 n  (h_z + r_z)/c}$, when evaluating the lower limit of the integral $k_{\rho} = 0$ in Eq.~\eqref{eq_scattered_dyadic_SI}.  Then, together with the exponential of the inverse Fourier transform of Eq.~\eqref{eq_elec_field_delta_dipole} we obtain an exponential of the form $\text{exp} \{ -i \omega_1 [t - \tau_1 - n   (h_z + r_z)/c ] \}$. Since in Eq.~\eqref{eq_elec_field_delta_dipole} the positions $\vec{h}(t)$ and $\vec{h}(\tau_1)$ are constants, we can identify $h_z = h(t)$ and $r_z = h(\tau_1)$ to conclude that the electric field $\vec{E}_{\text{aux}} ( \vec{h}(t), \vec{h}( \tau_1  ) , t - \tau_1 )$ is given in terms of Dirac deltas $\delta (t - \tau_1 - n |h (t) + h(\tau_1)|/c)$, and its derivatives because we have previously identified that it is the field generated by a frequency-independent dipole $\vec{p}_q$.\\

\begin{figure}[h!]
\includegraphics[scale=0.42]{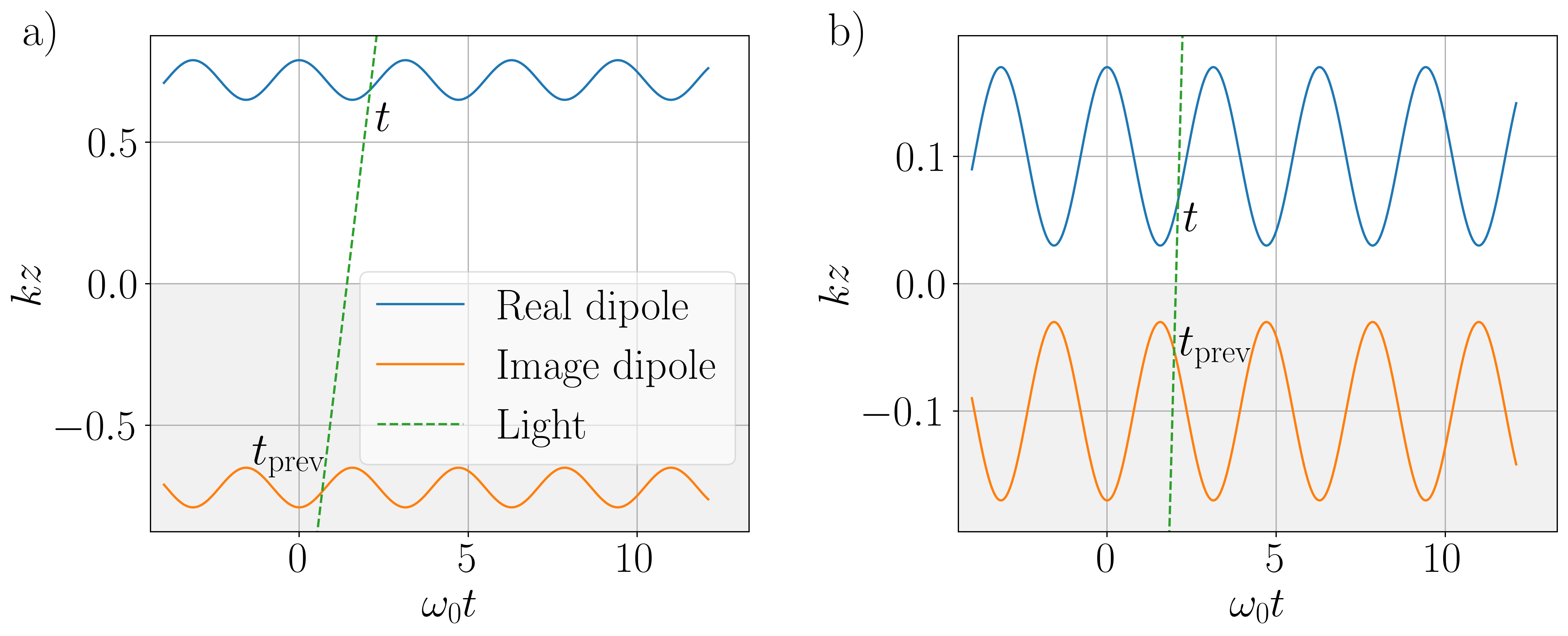}
\caption{(Color online) Graphical representation of the solution of Eq.~\eqref{eq_t_prev_SI} in a spacetime diagram, where \(k z\) denotes the coordinate perpendicular to the mirror scaled by the wave vector and \(\omega_0 t\) denotes time scaled by the dipole frequency. The upper and lower curves correspond to the trajectories of the real and image dipoles, respectively. The shaded region in the lower part represents the mirror. The dashed line represents the light trajectory (worldline), intersecting the dipole trajectories at times \(t\) (upper curve) and \(t_{\text{prev}}\) (lower curve). In a) the dipole moves at a larger distance while in b) the motion occurs very close to the mirror, in which case $t_{\text{prev}} \approx t$.}
\label{fig_panels_trajectories_compare}
\end{figure}

We can analyze the integral over $\tau_1$ in Eq.~\eqref{eq_es_field_supplementary}, given that the electric field $\vec{E}_{\text{aux}} ( \vec{h}(t), \vec{h}( \tau_1  ) , t - \tau_1 )$ is a sum of Dirac deltas $\delta (t - \tau_1 - n |h (t) + h(\tau_1)|/c )$ and its derivatives, weighted by factors that involve the effective distance $h(t) + h(\tau_1)$. In the general case, it is difficult to obtain an analytic expression for the electric field. However, the integrals that involve Dirac deltas composed with a function are straightforward to evaluate. For a general function $f(\tau_1)$, we can use the following result:
\begin{equation}\label{eq_int_Dirac_delta_SI}
\int_{- \infty}^{\infty} d \tau_1 \delta (g(\tau_1)) f(\tau_1) = \sum_{i} \frac{f(\tau_1^{(i)})}{|g'(\tau_1^{(i)})|},
\end{equation}
where $g(\tau_1) = t - \tau_1 - n |h (t) + h(\tau_1)|/c$ and the sum over $i$ corresponds to all the points $\tau_{1}^{(i)}$ that satisfy
\begin{equation}
g(\tau_1^{(i)}) = 0.
\end{equation}
In our case we have a single solution, which we will denote by $t_{\text{prev}}$, that satisfies
\begin{equation}\label{eq_t_prev_SI}
t_{\text{prev}} = t - n |h (t) + h(t_{\text{prev}})|/c.
\end{equation}
The last term corresponds to the addition of two distances, one from the dipole to the mirror $h(t_{\text{prev}})$ and the other one from the mirror to the dipole $h(t)$. This equation represents propagation of light from an image dipole inside a mirror, emitted at time $t_{\text{prev}}$ that arrives at time $t$. In a spacetime diagram, we can draw the two trajectories of the real and imaginary dipoles and connect two events by a lightlike path to obtain the graphical solution of Eq.~\eqref{eq_t_prev_SI}. In Fig.~\ref{fig_panels_trajectories_compare}, we illustrate the solutions for two cases: when the dipole oscillates about an average distance far from the mirror and when it is close to it. For small separations, the retardation becomes negligible, and we obtain $h(t_{\text{prev}}) \approx h(t)$, which will be useful for obtaining a Markovian dynamics. When performing the integrals that involve Dirac deltas, like in Eq.~\eqref{eq_int_Dirac_delta_SI}, we should neglect terms that involve the speed of the dipole divided by the speed of light $\dot{h}(\tau_1)/c$. The integral over $\tau_1$ in Eq.~\eqref{eq_es_field_supplementary} will allow us to obtain an electric field whose structure is identical to the case of a fixed-position dipole, the difference being that we have a time-dependent distance given by $|h(t) + h(t_{\text{prev}})|$, instead of a constant distance $2 h$.\\

We can use the Born approximation $\vec{p}(t) = \text{Re}[\vec{p}(0) e^{-i \omega_0 t} ]$ in the driving term $\vec{p}_{q} 2 \omega_0 \hbar^{-1} \vec{p}_{q} \cdot  \vec{E}_{s}(t)$ given in Eq.~\eqref{eq_es_field_supplementary} and follow the same procedure to obtain a decay rate and a Lamb shift, see Chapter 8 of Ref.~\cite{book_Novotny} and  Ref.~\cite{Chance_lifetime_emitting_molecule}. In this standard way, the Born approximation allows us to get $\vec{p}(t_{\text{prev}}) = \vec{p}(0) \text{exp}[-  i \omega_0 (t- n |h (t) + h(t_{\text{prev}})|/c)]$ and $\dot{\vec{p}}(t_{\text{prev}}) = -i \omega_0 \vec{p}(0) \text{exp}[-  i \omega_0 (t- n |h (t) + h(t_{\text{prev}})|/c)]$. Identifying the coefficients that define the dyadic Green's function, we find
\begin{equation}
\vec{p}_{q} \frac{2 \omega_0 \vec{p}_{q}}{\hbar} \cdot \vec{E}_s(t) \approx  \frac{q^2}{m} [ \mu \omega_0^2 \vec{n}_p \cdot \overline{\overline{G}}_{s}(\vec{h}(t),\vec{h}(t_{\text{prev}}), \omega_0) \cdot \vec{n}_p ] \vec{p}(0) e^{-i\omega_0 t}.
\end{equation}
When we consider the contribution from the free space dyadic Green's function $\overline{\overline{G}}_{0}(\vec{x},\vec{y}, \omega_0)$, we obtain a constant damping, as we will show below. We can approximate the imaginary terms as time derivatives of the dipole moment $-i \omega_0 \vec{p}(0)e^{-i \omega_0 t} \rightarrow \dot{\vec{p}}(t)$ to obtain the dynamical damping and Lamb shift,
\begin{equation}\label{eq_general_damping_lamb}
\begin{split}
\vec{p}_{q} \frac{2 \omega_0 \vec{p}_{q}}{\hbar} \cdot \vec{E} (t)  \approx \omega_0 \Delta \upsilon(\vec{h}(t),\vec{h}(t_{\text{prev}})) \vec{p}(t)  - \Gamma(\vec{h}(t),\vec{h}(t_{\text{prev}})) \frac{d}{dt} \vec{p}(t) ,
\end{split}
\end{equation}
where $\Gamma(\vec{h}(t),\vec{h}(t_{\text{prev}}))$ and $\Delta \upsilon(\vec{h}(t),\vec{h}(t_{\text{prev}}))$ are defined in the main text, see Eqs.~\eqref{eq_aux_decay_rates_imprf-nonMarkovian} and \eqref{eq_aux_Lamb_shift_imprf-nonMarkovian}. From the space-time diagrams shown in Fig.~\ref{fig_panels_trajectories_compare} we can see that at close distances $h(t_{\text{prev}}) \approx h(t)$ if we coarse grain the time with a time step $\Delta t$ that is larger than the time it would take the light to travel to the mirror and come back $\Delta t \gtrsim \max \{2 n h(t) /c\}$. In this short-distance regime we obtain the Markovian damping and Lamb shift given in Eqs.~\eqref{eq_aux_decay_rates_imprf} and \eqref{eq_aux_Lamb_shift_imprf}.\\

Let us address the contribution of the free space dyadic Green's function $\overline{\overline{G}}_{0}(\vec{x}, \vec{y} , \omega)$ in Eq.~\eqref{eq_driving_fields}. Following a similar procedure as in the case of the scattered field, we have
\begin{equation}\label{eq_0_e_field_supplementary}
\frac{\vec{p}_{q} 2 \omega_0 \vec{p}_{q} }{\hbar}  \cdot  \vec{E}_{0}(t) = \frac{1}{2\pi} \int_{- \infty}^{\infty}  d\omega_1   \int_{-\infty}^{\infty} d \tau_1   \;      e^{-i \omega_1  ( t - \tau_1) }     \Bigg( \frac{2 \omega_0 }{ \hbar } \Bigg) [ \mu_1 \omega_1^2]  \Bigg[   [\vec{p}_{q} \cdot  \overline{\overline{G}}_{0}(\vec{h}(t), \vec{h}( \tau_1  ) , \omega_1) \cdot \vec{p}_{q}]  \vec{p}(\tau_1)  \Bigg].
\end{equation}
We can find an explicit expression for the dyadic Green's function and perform the integral over $\omega_1$ \cite{book_Novotny}, to obtain an expression in terms of a virtual dipole $p_{\delta}(t)$,
\begin{equation}\label{eq_virt_dipole_Dirac}
p_{\delta}(t) = \int_{- \infty}^{\infty} d\omega_1 \; e^{-i \omega_1 t} p(\omega_1)  = p_q \int_{- \infty}^{\infty} d\omega_1 \; e^{-i \omega_1 t} = p_q 2 \pi \delta(t) .
\end{equation}
In analogy to Eq.~\eqref{eq_elec_field_delta_dipole} we can define an auxiliary electric field in terms of the free-space dyadic Green's function
\begin{equation}
\begin{split}
\vec{p}_{q} \cdot \vec{E}_{\text{aux}} ( \vec{h}(t), \vec{h}( \tau_1  ) , t - \tau_1) &= \vec{p}_{q} \cdot \int_{- \infty}^{\infty}  d\omega_1 \;      e^{-i \omega_1 (t - \tau_1) }   \mu_1 \omega_1^2         [  \overline{\overline{G}}_{0}(\vec{h}(t), \vec{h}( \tau_1  ) , \omega_1) \cdot \vec{p}_q] \\
&= p_q  \Bigg[ \frac{1 }{4 \pi \varepsilon_1} \Bigg(  \frac{2 \sqrt{\mu_1 \varepsilon_1}  \dot{p}_{\delta}(t - \tau_1 - t_{\text{ret}})}{R^2} + \frac{2 p_{\delta}(t - \tau_1 - t_{\text{ret}}) }{R^3} \Bigg)   \Bigg] ,
\end{split}
\end{equation}
where $R = |\vec{h}(t) - \vec{h}(\tau_1)|$ and $t_{\text{ret}} = n |\vec{h}(t) - \vec{h}(\tau_{1}) |/c$. When we integrate over $\tau_1$ we obtain an expression that involves the dipole $\vec{p}(t_{\text{prev}})$ evaluated at a previous time $t_{\text{prev}}$, which, differently from Eq.~\eqref{eq_t_prev_SI}, is given by
\begin{equation}
t_{\text{prev}} = t - n | \vec{h}(t) - \vec{h}(t_{\text{prev}}) | / c.
\end{equation}
One solution is $t_{\text{prev}} = t$, while other solutions would imply that the dipole moves at a speed comparable to the speed of light. Since we are in the nonrelativistic regime, we do not have such solutions and the driving field corresponds to the one at the instantaneous position of the dipole, as in the case where the position of the dipole is fixed. It is well known that the Born-Markov approximation allows us to obtain a constant damping. For instance, see Chapter 8 of Ref.~\cite{book_Novotny}. Hence, the free-space dyadic Green's function yields a constant damping $\gamma_n$.\\

Let us illustrate the calculation of the scattered electric field in Eq.~\eqref{eq_scattered_e_field_supplementary} for the case of a moving dipole in front of a perfect mirror. For a dipole moment that is perpendicular to the mirror $\vec{n}_p = \vec{e}_z$, and vectors $\vec{h}=(0,0,h_z)$ and $\vec{r}=(0,0,r_z)$, with $h_z,r_z > 0$, it is well known that in the static case the scattered electric field is given by \cite{book_Novotny}
\begin{equation}\label{eq_perp_orint_SI}
\mu_1 \omega_0^2 \; \vec{n}_p\cdot \overline{\overline{G}}_{s}(\vec{h}, \vec{r}, \omega_0) \cdot \vec{n}_p  = \frac{1}{4 \pi \varepsilon_1} \Bigg[  \frac{2 }{(h_{z} + r_{z})^3} - i \frac{2 \sqrt{\mu_1 \varepsilon_1} \omega_0 }{(h_{z} + r_{z})^2}  \Bigg] e^{ i  \sqrt{\mu_1 \varepsilon_1}\omega_0 (h_{z} + r_{z})}.
\end{equation}
For a parallel dipole $\vec{n}_p = \vec{e}_x$ or $\vec{n}_p = \vec{e}_y$, we have \cite{book_Novotny}
\begin{equation}\label{eq_parallel_orint_SI}
\mu_1 \omega_0^2 \; \vec{n}_p \cdot \overline{\overline{G}}_{s}(\vec{h}, \vec{r}, \omega_0) \cdot\vec{n}_p  = \frac{1}{4 \pi \varepsilon_1}  \Bigg[  \frac{1}{(h_{z} + r_{z})^3} - \frac{i \omega_0 \sqrt{\mu_1 \varepsilon_1}}{(h_{z} + r_{z})^2} - \frac{\omega_0^2 \mu_1 \varepsilon_1}{h_{z} + r_{z}}  \Bigg] e^{i   \sqrt{\mu_1 \varepsilon_1} \omega_0 (h_{z} + r_{z})}.
\end{equation}
Depending on the orientation of the dipole, we can substitute Eqs. \eqref{eq_perp_orint_SI} or \eqref{eq_parallel_orint_SI} into Eq.~\eqref{eq_es_field_supplementary} and then perform the integral over $\omega_1$. In Eq.~\eqref{eq_elec_field_delta_dipole}, we obtain
\begin{equation}\label{eq_aux_electric_field2}
\begin{split}
\vec{p}_{q} \cdot \vec{E}_{\text{aux}} ( \vec{h}(t), \vec{h}( \tau_1  ) , t - \tau_1) &= p_q  \Bigg[ \frac{1 }{4 \pi \varepsilon_1} \Bigg(  \frac{2 \sqrt{\mu_1 \varepsilon_1}  \dot{p}_{\delta}(t - \tau_1 - t_{\text{ret}})}{R^2} + \frac{2 p_{\delta}(t - \tau_1 - t_{\text{ret}}) }{R^3} \Bigg)   \Bigg] \\
&+ p_q   \Bigg[  \frac{1}{4 \pi \varepsilon_1} \Bigg(   \frac{\mu_1 \varepsilon_1 \ddot{p}_{\delta}(t - \tau_1 - t_{\text{ret}})}{R} + \frac{ \sqrt{\mu_1 \varepsilon_1} \dot{p}_{\delta}(t - \tau_1 - t_{\text{ret}}) }{R^2} + \frac{p_{\delta}(t - \tau_1 - t_{\text{ret}})}{R^3} \Bigg)  \Bigg] |\vec{n}_p \times \vec{e}_z| ,
\end{split}
\end{equation}
where we consider two cases for the orientation of the dipole $\vec{n}_p$, parallel or perpendicular to the mirror. For a general orientation of the dipole, we would have to consider a linear combination of the results obtained for the perpendicular and parallel cases, weighted by the orientation of the dipole. For instance, see Eq.~(10.26) in Ref.~\cite{book_Novotny}, where the decay of a dipole with arbitrary orientation is shown explicitly. We have defined $p_q = |\vec{p}_q|$, $R = |h(t) + h(\tau_1)|$, and $t_{\text{ret}}= \sqrt{\mu_1 \varepsilon_1} |h(t) + h(\tau_1)|$, where $h(t)$ is the predefined trajectory of the dipole, which occurs perpendicularly to the mirror. We have also introduced the virtual dipole $p_{\delta}(t)$ defined in Eq.~\eqref{eq_virt_dipole_Dirac}. When we substitute Eqs.~\eqref{eq_elec_field_delta_dipole} and \eqref{eq_aux_electric_field2} into Eq.~\eqref{eq_scattered_e_field_supplementary}, we can perform the integrals over $\tau_1$ since we have introduced the Dirac delta of $p_{\delta}(t - \tau_1 -t_{\text{ret}})$ and its derivatives. To evaluate these integrals, we can neglect the terms that involve the speed of the dipole divided by the speed of light, since we are in the nonrelativistic regime. As an example, we find the following integral
\begin{equation}
\begin{split}
I_1 &= \int_{-\infty}^{\infty} d \tau_1 \; \frac{d}{dx} \delta (x) \Bigg|_{x= t - \tau_1 - n |h(t) + h(\tau_1)|/c} \vec{p}(\tau_1)  \frac{1}{|h(t) + h( \tau_1  )|^2}\\
& \approx \int_{-\infty}^{\infty} d \tau_1\;   \delta (t - \tau_1 - n |h(t)+h(\tau_1)|/c) \Bigg[ -2 \frac{ \vec{p}(\tau_1)  }{|h(t)+h(\tau_1)|^3} \frac{d}{d \tau_1} h(\tau_1) + \frac{d}{d \tau_1} \vec{p}(\tau_1)  \frac{1}{|h(t)+h(\tau_1)|^2} \Bigg],
\end{split}
\end{equation}
where to integrate over $\tau_1$, we have to evaluate the zeros of the argument of the Dirac delta. For the perpendicular case $(\perp)$, Eq.~\eqref{eq_es_field_supplementary} becomes [see also Eq.~\eqref{eq_scattered_e_field_supplementary}]
\begin{equation}\label{eq_electric_field_moving_dip_perp}
\vec{p}_{q} \frac{2 \omega_0 \vec{p}_{q}}{\hbar} \cdot  \vec{E}_{s \perp} (t)  = \frac{q^2}{m} \frac{1}{4 \pi \varepsilon_1 } \Bigg[ \frac{2 \vec{p}(t_{\text{prev}})  }{[h(t)+h(t_{\text{prev}})]^3} + \frac{n}{c}\frac{2 \frac{d}{dt} \vec{p}(t_{\text{prev}})  }{[h(t)+h(t_{\text{prev}})]^2} - \frac{n}{c} \frac{2 \vec{p}(t_{\text{prev}})  }{|\vec{h}(t)+\vec{h}(t_{\text{prev}})|^3}  \dot{h}(t_{\text{prev}}) \Bigg],
\end{equation}
where $t_{\text{prev}}$ is determined by solving Eq.~\eqref{eq_t_prev_SI}. In a similar way, for the parallel case ($\parallel$) Eq.~\eqref{eq_es_field_supplementary} becomes [see also Eq.~\eqref{eq_scattered_e_field_supplementary}]
\begin{equation}\label{eq_electric_field_moving_dip_parallel}
\begin{split}
\vec{p}_{q} \frac{2 \omega_0 \vec{p}_{q}}{\hbar} \cdot \vec{E}_{s \parallel}(\vec{h}(t),t)  = \frac{q^2}{m} \frac{1}{4 \pi \varepsilon_1} \Bigg[ \frac{ \vec{p}(t_{\text{prev}})  }{[h(t)+h(t_{\text{prev}})]^3} + \frac{n}{c}\frac{ \frac{d}{dt} \vec{p}(t_{\text{prev}})  }{[h(t)+h(t_{\text{prev}})]^2} + \frac{n^2}{c^2} \frac{ \frac{d^2}{dt^2} \vec{p}(t_{\text{prev}})  }{|h(t) + h(t_{\text{prev}})|}   \Bigg],
\end{split}
\end{equation}
Using the image method, we can also obtain Eq.~\eqref{eq_electric_field_moving_dip_perp} (without the term proportional to $\dot{h}(t_{\text{prev}})$) and Eq.~\eqref{eq_electric_field_moving_dip_parallel} because they correspond to the electric field that was radiated by an image dipole at a distance $h(t)+h(t_{\text{prev}})$. The difference with the static case is the effective distance $h(t) + h(t_{\text{prev}})$, instead of $2 h$ in the static case. Therefore, we can use the Born approximation to obtain the non-Markovian damping and Lamb shift.

\section{Near field of the dipole close to an imperfect mirror}\label{subsec_elec_field_imperf}

From our analysis of the perfect mirror, we identify two key factors that enable the determination of instantaneous time-dependent decay rates and frequency shifts. The first is the proximity to the mirror, which permits neglecting retardation effects when the dipole motion is slow compared to the speed of light. The second is the weak-coupling regime. For the case of an imperfect mirror, we cannot calculate the reflected electric field explicitly. However, we can identify the terms that define the decay rate and the frequency shift close to the mirror. To this end, we extend the approach of Ref.~\cite{Kocharovsky_reversal_of_radiation_reaction}.

\subsubsection{Perpendicular dipole}
The electric field due to a dipole near the surface back onto itself is proportional to \cite{book_Novotny}:
\begin{equation}\label{eq_perp_imp_static_SI}
\mu_1 \omega_0^2 \vec{n}_p \cdot \overline{\overline{G}}_{s}(\vec{h}, \vec{h} , \omega_0) \cdot \vec{n}_p = \frac{i \mu_1 \omega_0^2 }{8 \pi k_1^2} \int_{0}^{\infty} dk_{\rho} \frac{k_\rho}{\sqrt{k_1^2 - k_\rho^2} }  2 k_\rho^2 r^p(k_\rho) e^{i \sqrt{k_1^2 - k_\rho^2} X},
\end{equation}
where $X=2h$, with $h$ the distance between the dipole and the mirror, and $\omega_0 n = c k_1$. We are interested in analyzing the close-distance regime of Eq.~\eqref{eq_perp_imp_static_SI}. First, we can analyze the Fresnel coefficient,
\begin{equation}
r^p = \frac{\varepsilon_2 \sqrt{ \frac{\omega^2_0}{c^2} \frac{\varepsilon_1 \mu_1 }{\varepsilon_0 \mu_0} - k_{\rho}^2 }  - \varepsilon_1 \sqrt{ \frac{\omega^2_0}{c^2} \frac{\varepsilon_2 \mu_2 }{\varepsilon_0 \mu_0}  - k_{\rho}^2} }{ \varepsilon_2 \sqrt{ \frac{\omega^2_0}{c^2} \frac{\varepsilon_1 \mu_1 }{\varepsilon_0 \mu_0} - k_{\rho}^2 }  + \varepsilon_1 \sqrt{ \frac{\omega^2_0}{c^2} \frac{\varepsilon_2 \mu_2 }{\varepsilon_0 \mu_0}  - k_{\rho}^2}  }.
\end{equation}
We can define the scaled variable $\kappa = k_{\rho} X$, which will serve as the new integration variable. Taking the limit $X \rightarrow 0$, corresponding to the short-distance regime near the mirror, we find
\begin{equation}\label{eq_approx_rp_SI}
r^p (k_\rho) \approx \frac{\varepsilon_2 \sqrt{-\kappa^2} - \varepsilon_1 \sqrt{-\kappa^2} }{ \varepsilon_2 \sqrt{-\kappa^2} + \varepsilon_1 \sqrt{-\kappa^2} } = \frac{ \varepsilon_2 - \varepsilon_1 }{ \varepsilon_2 + \varepsilon_1  }.
\end{equation}
In this limit, the following approximations hold:
\begin{equation}\label{eq_approx_exp_SI}
e^{i \sqrt{X^2k_1^2 - \kappa^2} } \approx e^{-\kappa} \;\; , \;\; \frac{ \kappa^3 }{\sqrt{X^2k_1^2 - \kappa^2} } \rightarrow \frac{\kappa^2 }{i}.
\end{equation}
Inserting Eqs.~\eqref{eq_approx_rp_SI} and \eqref{eq_approx_exp_SI} into Eq.~\eqref{eq_perp_imp_static_SI}, we get
\begin{equation}
\mu \omega_0^2 \vec{n}_p \cdot \overline{\overline{G}}(\vec{h}, \vec{h} , \omega_0) \cdot \vec{n}_p   \approx \frac{1}{4 \pi \varepsilon_1 X^3} \int_{0}^{\infty} d \kappa \; \kappa^2 r^p(\kappa) e^{-\kappa} \approx \frac{2}{4 \pi \varepsilon_1 X^3} \Bigg( \frac{ \varepsilon_2 - \varepsilon_1 }{ \varepsilon_2 + \varepsilon_1  }  \Bigg) ,
\end{equation}
from which we obtain the coefficients $\Delta \nu(\vec{h}(t))$ and $\gamma (\vec{h}(t))$ and therefore the coefficients in Eqs.~\eqref{eq_K_perp} and \eqref{eq_b_perp}. We leave the factor of two in the numerator to facilitate the comparison with the parallel case, as shown below in Eq.~\eqref{eq_approx_e-field_parallel_appendix}.

\subsubsection{Parallel dipole}
In the parallel case, the electric field is given by \cite{book_Novotny}
\begin{equation}\label{eq_field_parallel_dipole_imperf}
\begin{split}
\mu_1 \omega_0^2 \; \vec{n}_p\cdot \overline{\overline{G}}(\vec{h}, \vec{h}, \omega_0)\cdot \vec{n}_p =  \frac{i \mu_1 \omega_0^2}{8 \pi k_1^2} \int_{0}^{\infty} dk_{\rho} \frac{k_{\rho}}{ \sqrt{k_1^2 - k_\rho^2}  } 
[k_1^2 r^s - (k_1^2 - k_\rho^2) r^p] e^{ i \sqrt{k_1^2 - k_\rho^2}  X} .\\
\end{split}
\end{equation}
Following a similar analysis,
\begin{equation}
r^s = \frac{ \sqrt{ \frac{\omega^2_0}{c^2} \frac{\varepsilon_1 \mu_1 }{\varepsilon_0 \mu_0} - k_{\rho}^2 }  -  \sqrt{ \frac{\omega^2_0}{c^2} \frac{\varepsilon_2 \mu_2 }{\varepsilon_0 \mu_0}  - k_{\rho}^2} }{  \sqrt{ \frac{\omega^2_0}{c^2} \frac{\varepsilon_1 \mu_1 }{\varepsilon_0 \mu_0} - k_{\rho}^2 }  +  \sqrt{ \frac{\omega^2_0}{c^2} \frac{\varepsilon_2 \mu_2 }{\varepsilon_0 \mu_0}  - k_{\rho}^2}  }.
\end{equation}
Multiplying and dividing by $X$, we get
\begin{equation}
r^s = \frac{ \sqrt{X^2 \frac{\omega^2_0}{c^2} \frac{\varepsilon_1 \mu_1 }{\varepsilon_0 \mu_0} - X^2 k_{\rho}^2 }  -  \sqrt{ X^2  \frac{\omega^2_0}{c^2} \frac{\varepsilon_2 \mu_2 }{\varepsilon_0 \mu_0}  - X^2 k_{\rho}^2} }{  \sqrt{ X^2 \frac{\omega^2_0}{c^2} \frac{\varepsilon_1 \mu_1 }{\varepsilon_0 \mu_0} - X^2 k_{\rho}^2 }  +  \sqrt{ X^2 \frac{\omega^2_0}{c^2} \frac{\varepsilon_2 \mu_2 }{\varepsilon_0 \mu_0}  - X^2 k_{\rho}^2}  } \rightarrow \frac{\sqrt{-X^2 k_\rho^2} - \sqrt{-X^2 k_\rho^2}}{2 \sqrt{-X^2 k_\rho^2} } \approx 0.
\end{equation}
This shows that, in the near-field limit, the term involving $r^s$ can be safely neglected compared to the $r^p$ contribution. We can now integrate the remaining terms in Eq.~\eqref{eq_field_parallel_dipole_imperf} to find
\begin{equation}\label{eq_approx_e-field_parallel_appendix}
\begin{split}
\mu_1 \omega_0^2 \; \vec{n}_p\cdot \overline{\overline{G}}(\vec{h}, \vec{h}, \omega_0)\cdot  \vec{n}_p \approx &  \frac{i \mu_1 \omega_0^2 }{8 \pi k_1^2} \frac{1}{X^3} \int_{0}^{\infty} dk_{\rho} \; X  
[- \sqrt{X^2 k_1^2 - X^2 k_\rho^2} r^p] X k_{\rho} e^{ i \sqrt{X^2 k_1^2 - X^2 k_\rho^2}  } \\
 \approx & \frac{ \mu_1 \omega_0^2 }{8 \pi k_1^2} \frac{1}{X^3} r^p \int_{0}^{\infty} d\kappa \; \kappa^2
 e^{ -\kappa  } \\
\approx & \frac{1}{4 \pi \varepsilon_1 X^3} \Bigg( \frac{\varepsilon_2 - \varepsilon_1}{\varepsilon_2 + \varepsilon_1} \Bigg) .
\end{split}
\end{equation}
 Then, we can calculate the functions $\Delta \nu (\vec{h}(t))$ and $\gamma (\vec{h}(t))$ and therefore the coefficients in Eq.~\eqref{eq_Kb_parall}.

\section{Motion in front of an imperfect mirror}\label{subsec_motion_infront_mirror}
We will consider a modulation of the position given by $h(t) = R_0 [1 + \eta(t)]$. For a small amplitude $|\eta(t)| \ll 1$, we obtain
\begin{equation}\label{eq_small_variations_h_imperf}
\frac{1}{(2h(t))^3}  \approx  \frac{1}{8 R_0^3} [ 1 - 3  \eta (t) ].
\end{equation}
Hence, from Eq.~\eqref{eq_diff_eq_dipole_kb} we obtain:
\begin{equation}
\frac{d^2}{dt^2} \vec{p}(t)  + \omega_0^2 \vec{p}(t)  \approx \frac{K}{8 R_0^3}[ 1 - 3  \eta (t) ] \vec{p}(t)  - \Bigg( \frac{b}{8R_0^3}[ 1 - 3  \eta (t) ] + \gamma_{t} \Bigg) \frac{d}{dt} \vec{p}(t) ,
\end{equation}
where we are omitting the subscript of the perpendicular or parallel cases $(\perp, \parallel)$. Rearranging terms,
\begin{equation}
\frac{d^2}{dt^2} \vec{p}(t)  +    \Bigg( \omega_0^2 - \frac{K}{8 R_0^3}[ 1 - 3  \eta (t) ] \Bigg) \vec{p}(t)  + \Bigg( \frac{b}{8R_0^3}[ 1 - 3  \eta (t) ] + \gamma_{t} \Bigg) \frac{d}{dt}   \vec{p}(t)  =\vec{0}.
\end{equation}
This differential equation has the form of a parametric oscillator. Then, we can define a frequency function $\nu^2(t)$ and a damping function $\beta(t)$,
\begin{subequations}
\begin{align}
\nu^2 (t) &= \Bigg( \omega_0^2 - \frac{K}{8 R_0^3}[ 1 - 3  \eta (t) ] \Bigg),\\
\beta (t) &= \Bigg( \frac{b}{8R_0^3}[ 1 - 3  \eta (t) ] + \gamma_{t} \Bigg).
\end{align}
\end{subequations}
We can use a canonical transformation to remove the damping factor and obtain a differential equation of a parametric oscillator without damping. Let us define the expectation value of the operator $\vec{r}(t)$ as
\begin{equation}
\vec{p}(t)  = \text{exp} \Bigg[ - \int_0^{t} d s\; \frac{\beta (s)}{2}  \Bigg] \vec{r}(t) .
\end{equation}
Then, the expectation value $\vec{r}(t) $ obeys the differential equation
\begin{equation}
\frac{d^2}{dt^2} \vec{r}(t)  + \lambda^2(t) \vec{r}(t) = 0,
\end{equation}
where
\begin{equation}
\begin{split}
\lambda^2(t) &= \nu^2(t) - \frac{\dot{\beta}(t)}{2} - \frac{\beta^2(t)}{4}.\\
\end{split}
\end{equation}
In order to identify terms that can be neglected, we can scale the time variable with the frequency of the dipole $\omega_0$. Defining $\tau = \omega_0 t$, we obtain
\begin{equation}\label{eq_frequency_Gammat}
\begin{split}
\frac{\lambda^2 (\tau)}{\omega_0^2} &= \Bigg( 1 - \frac{K}{8 \omega_0^2 R_0^3}[ 1 - 3  \eta (\tau) ] \Bigg) + \frac{3 b \dot{\eta}(\tau)}{16 \omega_0 R_0^3}\\
& \approx \Bigg( 1 + \frac{3   K \eta (\tau)}{8 \omega_0^2 R_0^3} \Bigg) + \frac{3 b \dot{\eta}(\tau)}{16 \omega_0 R_0^3}\\
&= 1 +  c_{K} \eta (\tau) + m \dot{\eta}(\tau).\\
\end{split}
\end{equation}
We have omitted the constant term $K/(8\omega_0^2 R_0^3)$, which corresponds to a small static correction to the resonance frequency. The omitted factor is proportional to $\gamma_n / \omega_0$, which is a small quantity, that can be of the order of $10^{-5}$ as in the work by Kocharovsky et al.~\cite{Kocharovsky_reversal_of_radiation_reaction} or of the order $10^{-13}$ as in the original work by Drexhage with $\text{Eu}^{3+}$ ions \cite{Drexhage_influence_surface}. If the distance to the mirror is of the order of $\omega_0 n R / c \approx 0.1$ then we have a factor of $ (\omega_0 n R / c)^3 \approx 10^{3} $, and, as shown in Fig.~\ref{fig_real_imag_R}, we have $\text{Re}[R(\varepsilon_2 / \varepsilon_1)] \approx 10^{0}$. Then, the factor that we are omitting is smaller than one. We want the right hand side of Eq.~\eqref{eq_frequency_Gammat} to be equal to $1 + 2 \alpha \text{cos}(\tilde{\Omega} \tau)$. Hence, we have to solve a first-order inhomogeneous differential equation:
\begin{equation}\label{eq_diff_eq_eta_imprf_motion}
c_{K} \eta (\tau) + m \dot{\eta}(\tau) = 2 \alpha \text{cos}(\tilde{\Omega} \tau),
\end{equation}
which can be readily solved,
\begin{equation}
\label{eq:soleta}
\eta (\tau) = \eta_0 e^{-c_{K} \tau /m} + 2 \alpha \frac{( c_{K} \; \text{cos}(2 \tau) + 2 m \; \text{sin}(2 \tau)  )}{c_{K}^2 + 4 m^2}.
\end{equation}
The first term corresponds to a shift in the position from the mirror whose magnitude decreases exponentially in time so that at large times the average position of the dipole is $R_0$; it is due to the imperfections of the mirror $m \neq 0$. Then, once the transients have decayed, after a time $t > m/c_{K}$, the amplitude of the dipole is given by \footnote{If we keep the constant term $-K/(8 \omega_0^2 R_0^3)$ in the condition for amplification, we would have $\alpha (1- K/(8 \omega_0^2 R_0^3) )/2 \approx \alpha/2 $ since the product $\alpha K/(8 \omega_0^2 R_0^3)$ is small.
}
\begin{equation}\label{eq_behavior_dipole_exp}
\vec{p}(\tau)  \propto \text{exp} \Bigg[ \frac{1}{2} \Bigg( \alpha - \frac{b}{8 \omega_0 R_0^3} - \frac{\gamma_{t}}{\omega_0}  \Bigg) \tau   \Bigg].
\end{equation}
To obtain amplification, we need the argument of the exponential to be positive. Since we are restricted to small modulations, we have
\begin{equation}
|\eta_0| \ll 1.
\end{equation}
We are also considering values of the distance and permittivities such that
\begin{equation}
\frac{\gamma_{t}}{\omega_0} < c_{K} \ll 1,
\end{equation}
and
\begin{equation}
\frac{\gamma_{t}}{\omega_0} < m = \frac{3 b }{16 \omega_0 R_0^3} \ll 1.
\end{equation}
Then, we can choose an amplitude of modulation $\alpha$ such that we can obtain an increase in the amplitude of the dipole
\begin{equation}
\alpha > \frac{b}{8 \omega_0 R_0^3} + \frac{\gamma_{t}}{\omega_0}.
\end{equation}
However, we also need to make sure that $\alpha$ should be small so that $\eta (\tau)\ll 1$. 
More precisely, notice that we can find a real valued phase $\delta$ such that,
\begin{equation}
\frac{( c_{K} \; \text{cos}(\tilde{\Omega} \tau) + \tilde{\Omega} m \; \text{sin}(\tilde{\Omega} \tau)  )}{ \sqrt{ c_{K}^2 + (\tilde{\Omega} m)^2 }} = \text{cos}(\delta) \text{cos}(\tilde{\Omega} \tau) + \text{sin}(\delta)\text{sin}(\tilde{\Omega} \tau)  = \text{cos}(\tilde{\Omega} \tau - \delta),
\end{equation}
where the phase $\delta$ is given by
\begin{equation}
\text{tan}(\delta) = \frac{\tilde{\Omega} m}{c_{K}}.
\end{equation}
Hence, after transients have decayed, Eq.~\eqref{eq:soleta} reads
\begin{equation}
\eta(\tau) \approx  2 \alpha \frac{( c_{K} \; \text{cos}(\tilde{\Omega} \tau) + \tilde{\Omega} m \; \text{sin}(\tilde{\Omega} \tau)  )}{  c_{K}^2 + (\tilde{\Omega} m)^2 }  = \frac{2 \alpha}{\sqrt{ c_{K}^2 + (\tilde{\Omega} m)^2 }} \text{cos}(\tilde{\Omega} \tau - \delta)<1,
\end{equation}
which can be fulfilled by
\begin{equation}
\frac{2 \alpha}{\sqrt{ c_{K}^2 + (\tilde{\Omega} m)^2 }} < 1.
\end{equation}
We can use the definition of $c_K$ and $m$ to translate this condition into
\begin{equation}
2 \alpha < \sqrt{ \Bigg(\frac{3   K }{8 \omega_0^2 R_0^3}\Bigg)^2 + \Bigg( \tilde{\Omega} \frac{3 b }{16 \omega_0 R_0^3} \Bigg)^2 }.
\end{equation}
For the case $\tilde{\Omega}=2$ (which is known to lead to amplification \cite{Kovacic_Mathieu_eq}), we find that $\alpha$ should obey the following inequalities:
\begin{equation}
\frac{m}{3} + \frac{\gamma_{t}}{\omega_0} < \alpha < \frac{1}{2} \sqrt{c_{K}^2 + 4 m^2 } .
\end{equation}
Then, we should choose the distance and permittivities such that the following strict inequality holds
\begin{equation}
- \frac{5}{9}m^2 + \Bigg( \frac{\gamma_t}{\omega_0} \Bigg)^2 + \frac{4 m \gamma_t}{3 \omega_0} < \frac{c_{K}^2 }{4}.
\end{equation}
We can recover the case of a perpendicular dipole in front of a perfect mirror when $b \rightarrow 0$ and $\gamma_{t} = 2 \gamma_n$. While in Eq.~\eqref{eq_K_perp} we obtain 
\begin{equation}
K_{\text{perf} \perp} = \frac{q^2}{m} \frac{2}{4 \pi \varepsilon_1} = \frac{3 \gamma_n \omega_0}{k^3},
\end{equation}
where we used
\begin{equation}
\frac{q^2}{m \omega_0} = \frac{\gamma_n 6 \pi \varepsilon_1}{k^3}.
\end{equation}
In this case, Eq.~\eqref{eq_diff_eq_eta_imprf_motion} can be solved algebraically for $\eta(\tau)$. Hence, we can identify an amplitude of modulation for $\eta(\tau) = A \text{cos}(\tilde{\Omega}\tau)$ given by 
\begin{equation}
\alpha = \frac{3 K A}{16 \omega_0^2 R_0^3},
\end{equation}
which corresponds to Eq.~\eqref{eq_alpha_perf_mirr}.

\section{Threshold amplitude of modulation for different distances}

\begin{figure}[h!]
\includegraphics[scale=0.42]{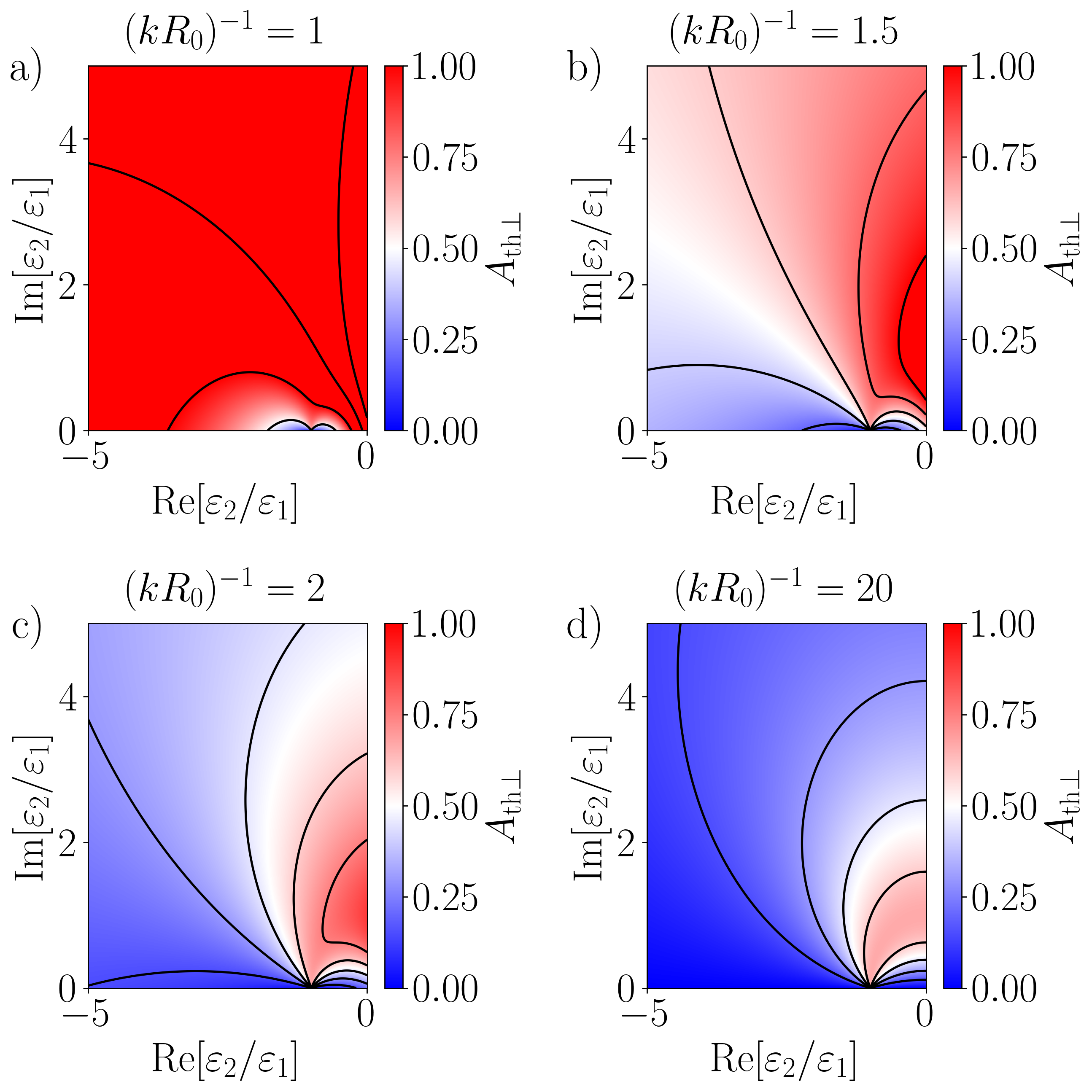}
\caption{(Color online) Threshold amplitude of modulation $A_{\mathrm{th} \perp}$ required to achieve amplification for a perpendicular dipole at different distances near a lossy metal, as given by Eq.~\eqref{eq_thresh_amp_LM}. Results are shown for distances corresponding to a) $(kR_0)^{-1}=1$, b) $(kR_0)^{-1}=1.5$, c) $(kR_0)^{-1}=2$, and d) $(kR_0)^{-1}=20$.}
\label{fig_panels_thresh_dis}
\end{figure}

Here, we present the behavior for the threshold amplitudes of modulation required to obtain amplification for distances slightly away from the mirror, as given in Eq.~\eqref{eq_thresh_amp_LM}. In Fig.~\ref{fig_panels_thresh_dis}, we can see that as we decrease the distance to the mirror, amplification becomes possible in a larger parameter regime. Also, we can observe convergence of the behavior at close distances when comparing Fig.~\ref{fig_panels_thresh_dis} d) with Fig.~\ref{fig_thresh_ampl_LM}.

\end{document}